\documentclass[twocolumn]{aastex63}
\usepackage{amsmath}

\received{}
\revised{}
\accepted{}

\shorttitle{Molecular Wind in NGC 1482}
\shortauthors{Salak et al.}

\begin{document}

\title{Molecular Gas Outflow in the Starburst Galaxy NGC 1482}

\correspondingauthor{Dragan Salak}
\email{salak.dragan.fm@u.tsukuba.ac.jp}

\author[0000-0002-3848-1757]{Dragan Salak}
\affiliation{Tomonaga Center for the History of the Universe, University of Tsukuba, 1-1-1 Tennodai, Tsukuba, Ibaraki 305-8571, Japan}

\author[0000-0002-5461-6359]{Naomasa Nakai}
\affiliation{School of Science and Technology, Kwansei Gakuin University, 2-1 Gakuen, Sanda, Hyogo 669-1337, Japan}

\author[0000-0003-1420-4293]{Kazuo Sorai}
\affiliation{Tomonaga Center for the History of the Universe, University of Tsukuba, 1-1-1 Tennodai, Tsukuba, Ibaraki 305-8571, Japan}
\affiliation{Department of Physics, Faculty of Science, Hokkaido University, Kita 10 Nishi 8, Kita-ku, Sapporo, Hokkaido 060-0810, Japan}
\affiliation{Department of Cosmosciences, Graduate School of Science, Hokkaido University, Kita 10 Nishi 8, Kita-ku, Sapporo, Hokkaido 060-0810, Japan}
\affiliation{Graduate School of Pure and Applied Sciences, University of Tsukuba, 1-1-1 Tennodai, Tsukuba, Ibaraki 305-8571, Japan}

\author[0000-0002-7616-7427]{Yusuke Miyamoto}
\affiliation{National Astronomical Observatory of Japan, 2-21-1 Osawa, Mitaka, Tokyo 181-8588, Japan}

\begin{abstract}

Galactic winds are essential to regulation of star formation in galaxies. To study the distribution and dynamics of molecular gas in a wind, we imaged the nearby starburst galaxy NGC 1482 in CO (\(J=1\rightarrow0\)) at a resolution of \(1\arcsec\) (\(\approx100\) pc) using the Atacama Large Millimeter/submillimeter Array. Molecular gas is detected in a nearly edge-on disk with a radius of 3 kpc and a biconical outflow emerging from the central 1 kpc starburst and extending to at least 1.5 kpc perpendicular to the disk. In the outflow, CO gas is distributed approximately as a cylindrically symmetrical envelope surrounding the warm and hot ionized gas traced by H\(\alpha\) and soft X-rays. The velocity, mass outflow rate, and kinetic energy of the molecular outflow are \(v_\mathrm{w}\sim100~\mathrm{km~s^{-1}}\), \(\dot{M}_\mathrm{w}\sim7~M_\sun~\mathrm{yr}^{-1}\), and \(E_\mathrm{w}\sim7\times10^{54}~\mathrm{erg}\), respectively. \(\dot{M}_\mathrm{w}\) is comparable to the star formation rate (\(\dot{M}_\mathrm{w}/\mathrm{SFR}\sim2\)) and \(E_\mathrm{w}\) is \(\sim1\%\) of the total energy released by stellar feedback in the past \(1\times10^7~\mathrm{yr}\), which is the dynamical timescale of the outflow. The results indicate that the wind is starburst driven.

\end{abstract}


\section{Introduction} \label{sec:intro}

Galactic winds (superwinds) are outflows of gas and dust from the central regions of galaxies. Driven by feedback from vigorous star formation and/or jets and radiation from an active galactic nucleus (AGN), the winds can transport metal-enriched interstellar material into the circumgalactic (CGM) and intergalactic medium (IGM), thereby regulating star formation and the growth of the central supermassive black hole (e.g., \citealt{VCB05,MQT05,HQM12,Rup19}). The feedback processes are thus critical in galaxy evolution throughout cosmic history and motivate a large number of observational and theoretical studies [e.g., see reviews in \citet{Rup18}, \citet{Zha18}, and \citet{Vei20}, and the references therein].

Outflows are comprised of multiple phases of the interstellar medium (ISM), including ionized and neutral gases, dust grains, and polycyclic aromatic hydrocarbons (PAHs), and understanding their properties is challenging because of the need for multi-wavelength observations with sensitivity to extended, low-brightness structure. In particular, molecular gas outflows driven by stellar feedback have been a subject of intensive research in recent years owing to the essential role of H\(_2\) gas in star formation. Molecular outflows have been observed at high resolution and studied in detail in several starburst galaxies in the local (\(D\lesssim 20\) Mpc) universe, including M82 (e.g., \citealt{WWS02,Sal13,Ler15}), NGC 253 (e.g., \citealt{Bol13,Zsc18,Kri19}), NGC 1808 (e.g., \citealt{Sal16,Sal17,Sal18}), NGC 2146 \citep{Tsa09}, and NGC 3628 \citep{Tsa12}. These and other studies suggest that molecular gas outflows carry a significant fraction of the mass and kinetic energy of starburst-driven winds and exhibit mass outflow rates comparable to or larger than the star formation rates (SFR) in regions where they originate, thereby regulating star formation. However, compared to the number of nearby galaxies where ionized gas outflows have been observed and studied in detail (e.g., \citealt{SB10,Hec15}), high-resolution imaging of molecular winds at mm/sub-mm wavelengths is still limited. Increasing the number of high-resolution, multi-wavelength observations of nearby systems, as an important step to grasp the full picture, has now become feasible with modern facilities. Such studies are important not only to probe the energetics of molecular winds and their negative feedback on star formation, but also to understand the physical processes taking place inside the winds. Nearby galaxies are currently the only laboratories where we can achieve cloud-scale angular resolution to investigate the morphology and dynamics of each ISM phase in the wind. In this paper, we report on the first 100-pc resolution observations of molecular gas in the nearby galaxy NGC 1482 traced by the rotational line \(^{12}\)CO (\(J=1\rightarrow0\)), hereafter CO (1--0). The \(J=1\rightarrow0\) transition to the ground state is a particularly useful probe of cold outflows because it can trace the low-density (\(n_\mathrm{H_2}\sim10^{2\mathrm{-}3}~\mathrm{cm^{-3}}\)) molecular medium.

\begin{figure}
 \centering
  \includegraphics[width=0.475\textwidth]{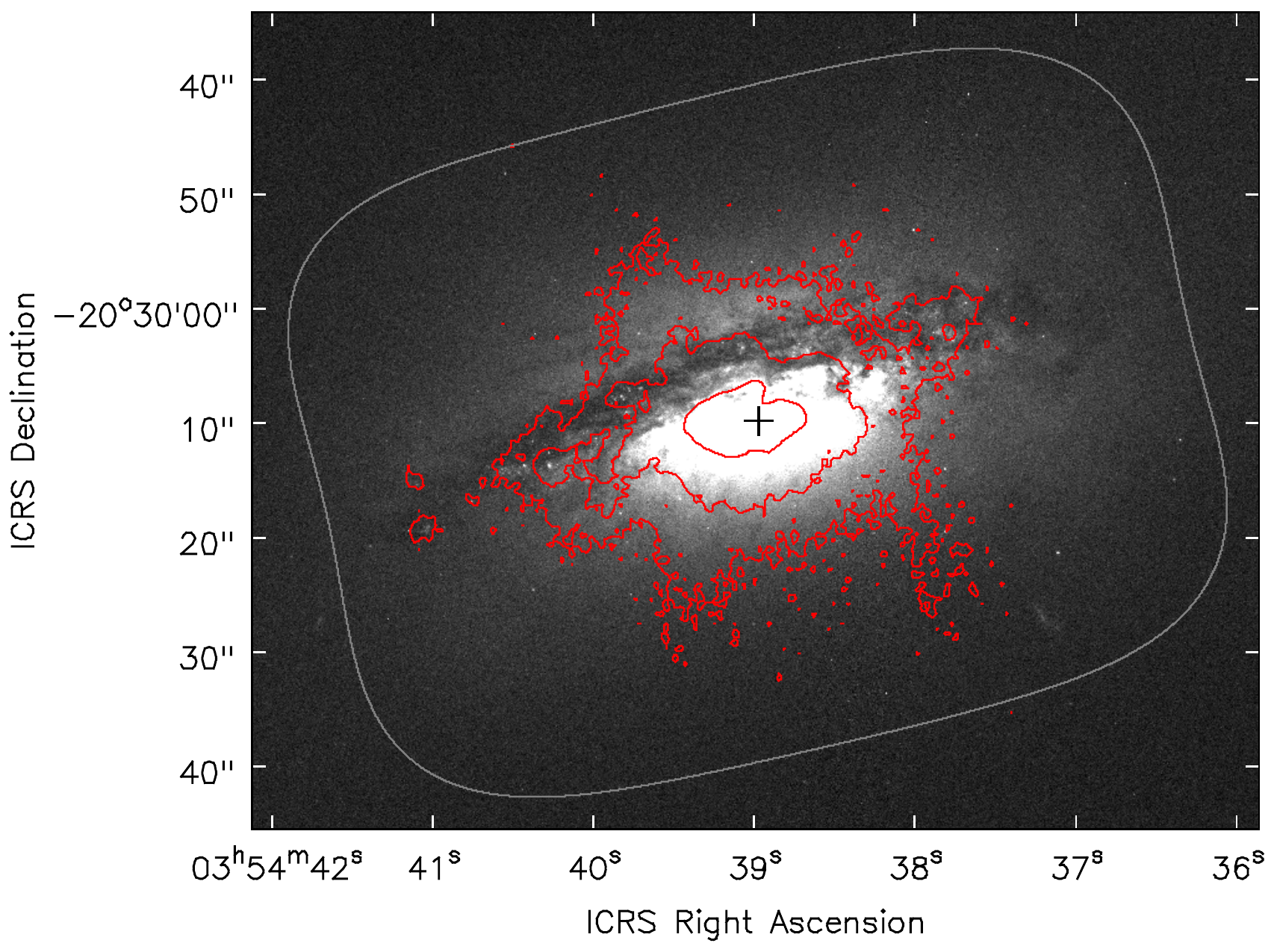}
 \caption{\(R\)-band (F621M) HST image of NGC 1482 from Hubble Legacy Archive with
H\(\alpha\) contours (red) at \((0.001,0.01,0.1)\times I_\mathrm{max}\) (maximum intensity) from \citet{Ken03}. The ALMA mosaic center is indicated by ``+'' and the primary beam response is plotted at 80\% as a gray contour.\label{fig:gal}}
\end{figure}

NGC 1482 is a nearby (\(D=19.6\) Mpc; \citealt{Tul88}), early-type galaxy in the Eridanus group of galaxies \citep{Fou92}. In Figure \ref{fig:gal}, an optical image acquired by the \emph{Hubble Space Telescope} (HST) shows a central bulge surrounded by a nearly edge-on gaseous disk (dust lane), and H\(\alpha\) contours reveal warm ionized gas extending perpendicular to the disk. The galaxy is well known for a biconical, hourglass-shaped superwind clearly visible in H\(\alpha\) and soft X-rays, believed to be powered in the underlying starburst disk \citep{HD99,VR02,Str04a}. The fact that the outflow is nearly perpendicular to the nearly edge-on gaseous disk makes this galaxy one of the valuable targets to study the outflow morphology. Extended structure was also detected in other tracers of ionized gas, such as [\ion{N}{2}], [\ion{S}{2}], [\ion{S}{3}], and [\ion{O}{3}] \citep{SB10,Zas13}. \citet{VR02}, \citet{SB10}, and \citet{Zas13} analyzed the optical emission lines and found evidence of shock ionization in the extraplanar ionized gas with high values of [\ion{N}{2}]\(\lambda\)6585/H\(\alpha\) and [\ion{S}{2}]\(\lambda\)6718/H\(\alpha\) intensity ratios. The wind is also glowing in soft X-rays extending up to 5 kpc from the starburst nucleus \citep{Str04a}. \citet{VR02} reported an outflow velocity of 250 km s\(^{-1}\) for the warm ionized gas, consistent with typical velocities of starburst-driven winds. The basic properties of the galaxy are summarized in Table \ref{tab:gal}.

\begin{table}
\begin{center}
\caption{Basic Parameters of NGC 1482}\label{tab:gal}
\begin{tabular}{llc}
\tableline\tableline
Parameter & Value & Reference \\
\tableline
Morphology & SA0\(^{+}\) pec & (1) \\
Central activity & \ion{H}{2} & (2) \\
Distance & 19.6 Mpc (\(1\arcsec=95\) pc) & (3) \\
\((\alpha,\delta)_\mathrm{ICRS}^\mathrm{CO}\) &
\(\mathrm{3^h54^m38\fs917},-20\arcdeg30\arcmin07\farcs67\) & (4) \\
\((\alpha,\delta)_\mathrm{ICRS}^{K_S}\) &
 \(\mathrm{3^h54^m38\fs933},-20\arcdeg30\arcmin08\farcs08\) & (4) \\
\(v_\mathrm{sys}\) (LSR) & \(1831\pm5\) km s\(^{-1}\) & (4) \\
Inclination & \(76\fdg1\pm2\fdg4\) & (4) \\
Position angle & \(-74\fdg35\pm0\fdg43\) & (4) \\
\tableline
\end{tabular}
\end{center}
\tablerefs{(1) \citet{deV91}, (2) NED classification, (3) \citet{Tul88}, (4) This work. The systemic velocity \(v_\mathrm{sys}\) and inclination were derived from CO data. The position angle is from a \(K_S\)-band image.}
\end{table}

Molecular gas was recently imaged in CO (1--0) at a resolution of \(17\arcsec\) by the Nobeyama 45-m telescope, as part of the CO Multi-line Imaging of Nearby Galaxies (COMING) project \citep{Sor19}. The central region of the galaxy was found to be CO-luminous (line luminosity \(L'_\mathrm{CO}=5.35\times10^8~\mathrm{K~km~s^{-1}~pc^2}\)), indicating the presence of a large molecular gas reservoir in the central few kiloparsecs and a possibility that molecular gas is entrained in the superwind. To study the distribution and dynamics of molecular gas, as the key ingredient in NGC 1482's starburst activity, we have carried out high-resolution CO (1--0) observations using the Atacama Large Millimeter/submillimeter Array (ALMA).

This paper is organized as follows. In section \ref{sec:obs}, we describe the ALMA observations and data reduction. The results are presented in \ref{sec:res}, where we show the distribution of molecular gas, traced by CO (1--0), and star-forming nucleus, revealed by the 100 GHz continuum. In section \ref{sec:gd}, we model the molecular gas disk and derive its basic geometric and kinematic parameters. We present evidence of a molecular gas outflow from the disk and analyze its dynamics in section \ref{sec:out}. The driving mechanism of the outflow and the origin and fate and the ejected material are discussed in section \ref{sec:dis}. The discussion is followed by a summary in section \ref{sec:sum}.

\section{Observations}\label{sec:obs}

The observations were carried out by ALMA in cycle 7 using 43 antennas of the 12-m array (TM) and 10 antennas of the 7-m Atacama Compact Array (ACA). Four spectral windows were tuned in band 3 to include the CO (1--0) line at the rest frequency of \(\nu_\mathrm{rest}=115.2712018\) GHz and the 100 GHz continuum. In order to cover a wide field of the central radius \(r\approx4\) kpc that includes the starburst and superwind regions, the galaxy was imaged over a rectangular mosaic of \(80\arcsec\times60\arcsec\) at a position angle of \(-77\arcdeg\) adopted from \citet{Sor19}, where the negative value is measured from north to west. This position angle is equivalent to \(283\arcdeg\) measured from north to east up to the receding side of the galactic disk, which is redshifted with respect to the systemic velocity (section \ref{sec:kin}). The region covered by the mosaic is indicated in Figure \ref{fig:gal}. Since superwinds are generally low-brightness, extended structures, it was essential to recover large-scale emission using ACA with its short baselines that achieved a maximum recoverable scale of 58\arcsec (\(\approx5.5\) kpc). Single-dish total power observations were not conducted, because the maximum recoverable scale of ACA was larger than the expected target size. However, as shown in section \ref{sec:mgd} below, the total flux of the ALMA map is consistent with that obtained by the Nobeyama 45-m telescope, indicating that the flux is recovered by the interferometer (TM combined with ACA) observations. The mosaic was centered at  \((\alpha,\delta)_\mathrm{ICRS}^{\mathrm{mosaic}}=(\mathrm{3^h54^m38\fs965,-20\arcdeg30\arcmin09\farcs650})\), which was adopted from NASA/IPAC Extragalactic Database (Figure \ref{fig:gal}). A summary of observations is given in Table \ref{tab:obs}.

\begin{table}
\begin{center}
\caption{Observational Summary}\label{tab:obs}
\begin{tabular}{lcc}
\tableline\tableline
Parameter & TM & ACA \\
\tableline
Observation date & 2020 Mar 3, 4 & 2019 Nov 23, Dec 2 \\
No. of antennas & 43 & 9, 10 \\
No. of pointings & 11 & 3  \\
Baselines (m) & 15-783 & 9-45  \\
On source time (minutes) & 27.7 & 63.5 \\
Flux, bandpass calibrator & J0334-4008 & J0423-0120 \\
Phase calibrator & J0349-2102 & J0348-1610  \\
\tableline
\end{tabular}
\end{center}
\end{table}

\begin{table}
\begin{center}
\caption{Image Parameters}\label{tab:imp}
\begin{tabular}{lcc}
\tableline\tableline
Parameter & Continuum & CO (1--0) Cube \\
\tableline
Weighting & Briggs & natural \\
FWHM \(b_\mathrm{maj}\) (arcsec) & 1.379 & 1.559 \\
FWHM \(b_\mathrm{min}\) (arcsec) & 0.949 & 1.083 \\
Beam position angle (degree) & -64.8 & -61.5 \\
Total bandwidth (GHz) & 6.7 & ... \\
Spectral resolution (km s\(^{-1}\)) & ... & 10.16 \\
Sensitivity rms (mJy beam\(^{-1}\)) & 0.031 & 2.1 \\
\tableline
\end{tabular}
\end{center}
\end{table}

Data reduction was conducted using the Common Astronomy Software Applications (CASA) package \citep{McM07}. The TM and ACA data were separately calibrated using the CASA (version 5.6.1-8) pipeline and then split into line (CO) and continuum measurement sets.

The calibrated CO visibilities were first continuum-subtracted using the CASA task \emph{uvcontsub}. The combined TM and ACA visibilities were then deconvolved together using the task \emph{tclean}. The image reconstruction was performed in mosaic mode with natural weighting and a spectral resolution of \(\Delta v=10.16\) km s\(^{-1}\). The full width at half maximum (FWHM) of the synthesized beam is \(b_\mathrm{maj}=1\farcs559\) (major axis) and \(b_\mathrm{min}=1\farcs083\) (minor axis) at a position angle of \(-61\fdg5\). For the adopted distance to the galaxy (19.6 Mpc), the linear scale is \(1\arcsec=95~\mathrm{pc}\), so the achieved angular resolution is \(\approx100~\mathrm{pc}\). The rms sensitivity at a spectral resolution of 10.16 km s\(^{-1}\) is 2.1 mJy beam\(^{-1}\), as calculated using the CASA task \emph{imstat} over emission-free channels.

The 100 GHz continuum visibilities were acquired from TM and ACA over four spectral windows with central sky frequencies of 100.6, 102.6, 112.7, and 114.6 GHz. First, the data were flagged to remove the contribution from CO (1--0) and CN (1--0) (doublet at 113 GHz) lines, and the total synthesized bandwidth after flagging was 6.7 GHz. The image was then reconstructed using the task \emph{tclean} in mosaic and multi-frequency synthesis modes with Briggs weighting (robust parameter 0.5). The FWHM beam size of the final image is \(b_\mathrm{maj}\times b_\mathrm{min}=1\farcs379\times0\farcs949\) at a position angle of \(-64\fdg8\). The rms sensitivity of the continuum image in a bandwidth of 6.7 GHz is 31 \(\mu\)Jy beam\(^{-1}\). The image parameters are summarized in Table \ref{tab:imp}.

Primary beam correction was applied to all ALMA data presented in this paper. The accuracy of flux calibration in ALMA band 3 is better than 5\%. The velocity is expressed in radio definition with respect to the local standard of rest (LSR).

\section{Results}\label{sec:res}

In this section, we present the images of CO (1--0) integrated intensity, velocity, and the distribution of 100 GHz continuum. We derive the molecular gas mass and star formation rate using the CO and continuum data, respectively.

\subsection{Molecular Gas Distribution}\label{sec:mgd}

\begin{figure*}
 \centering
  \includegraphics[width=0.8\textwidth]{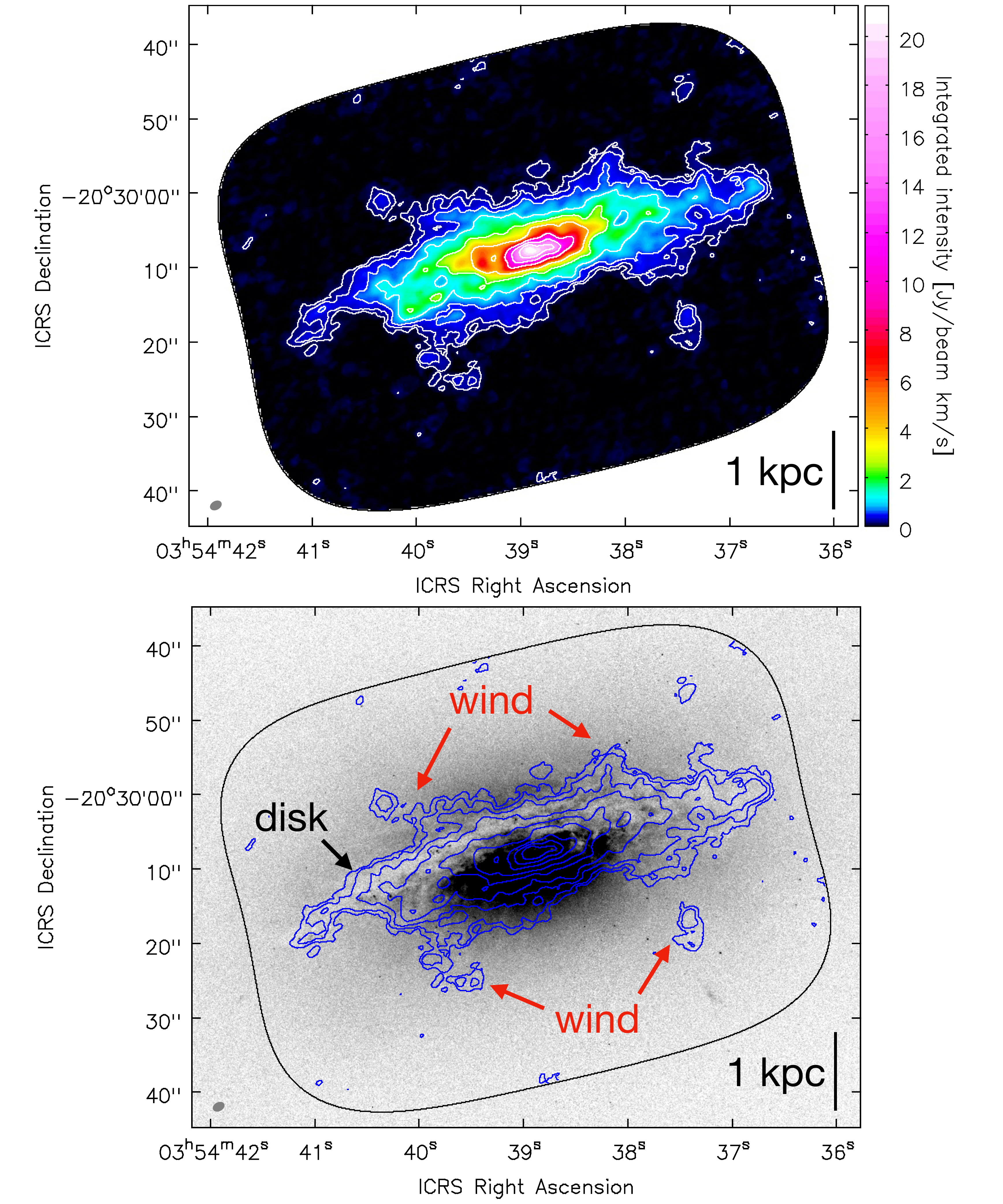}
 \caption{\emph{Top.} CO (1--0) integrated intensity image with contours plotted at \((0.00625,0.0125,0.025,0.05,0.1,0.2,0.4,0.6,0.8)\times21.3~\mathrm{Jy~beam^{-1}~km~s^{-1}}\) (maximum). The beam size (filled ellipse) is shown at the bottom left corner. \emph{Bottom.} CO (1--0) contours on a \(R\)-band HST negative image. The nearly edge-on molecular disk (dust lane) and extraplanar streamers are indicated.\label{fig:co}}
\end{figure*}

Figure \ref{fig:co} shows the CO (1--0) integrated intensity (moment 0) map, defined as \(I_\mathrm{CO}=\int \mathcal{S}_\mathrm{CO}\mathrm{d}v\), where \(\mathcal{S}_\mathrm{CO}\) is the intensity in Jy beam\(^{-1}\). To create the integrated intensity image, the data cube was first smoothed to twice the beam size with the CASA task \emph{imsmooth}, and the smoothed cube clipped at \(3\sigma\) was applied as a mask to the original cube. The masked cube at original resolution was integrated over velocity with \emph{immoments}.

A comparison with the optical image in Figure \ref{fig:co} (lower panel) shows that most of molecular gas is concentrated in a thin disk with a radius of 3 kpc that corresponds to the dust lane.  The CO distribution is concentrated in the central 700 pc region that coincides with the starburst nucleus. We determined the coordinates of the CO peak by fitting an elliptical Gaussian over a region twice the beam size centered at the maximum pixel value. This was performed using the CASA task \emph{imfit} and the result is given in Table \ref{fig:gal}. The absolute positional accuracy of the ALMA image is 5\% of the beam size (\(\approx0\farcs05\)).

The distribution also shows streamers, denoted by ``wind'' in Figure \ref{fig:co} (lower panel), that extend up to a height of 1.5 kpc symmetrically in four directions north and south of the disk. This gas is extraplanar and appears to be associated with the superwind observed in tracers of warm ionized gas. We will discuss the streamers in sections \ref{sec:out} and \ref{sec:dis}.

The total molecular gas mass including the contribution from helium and other elements can be obtained from the measured CO (1--0) integrated intensity using the formula

\begin{equation}\label{eq:mol}
M_\mathrm{mol}=1.05\times10^4(X_\mathrm{CO}/X_\mathrm{MW})D^2\int S_\mathrm{CO}\mathrm{d}v,
\end{equation}
where \(M_\mathrm{mol}\) is in \(M_\sun\), \(D\) is the distance to the galaxy in Mpc, \(S_\mathrm{CO}\) is the flux density in Jy, \(v\) is the velocity in km s\(^{-1}\) over which the flux density is integrated, and \(X_\mathrm{CO}\) is the CO-to-H\(_2\) conversion factor. The conversion factor is normalized relative to the standard value of \(X_\mathrm{MW}=2.0\times10^{20}~\mathrm{cm^{-2}(K~km~s^{-1})^{-1}}\) for the Galactic disk \citep{BWL13}. The total integrated flux density within the region in Figure \ref{fig:co} is \(\int S_\mathrm{CO}\mathrm{d}v=677~\mathrm{Jy~km~s^{-1}}\), and applying the Galactic conversion factor \(X_\mathrm{CO}=X_\mathrm{MW}\), the total molecular gas mass in this region is \(M_\mathrm{mol}=2.7\times10^9~M_\sun\). This is consistent within 15\% with \(2.3\times10^9~M_\sun\) in \citet{Sor19}, whose map covers a slightly smaller region at a resolution of \(17\arcsec\). On the other hand, if we apply a lower conversion factor of  \(X_\mathrm{CO}=0.5\times10^{20}~\mathrm{cm^{-2}(K~km~s^{-1})^{-1}}\), which is recommended for the Galactic center region (\(r<500\) pc; \citealt{BWL13}), the value is \(6.8\times10^8~M_\sun\). Since the starburst region in NGC 1482 is within \(r\approx7\arcsec\) (\(\approx760\) pc), which is small compared to the 3 kpc disk, the larger conversion factor is probably more appropriate for the outer part of the molecular gas disk. By comparison, the mass of atomic gas \ion{H}{1} in the galaxy is reported to be \(M_\mathrm{HI}=(1.35\pm0.15)\times10^9~M_\sun\) \citep{HS05}.

\begin{figure}
 \centering
  \includegraphics[width=0.475\textwidth]{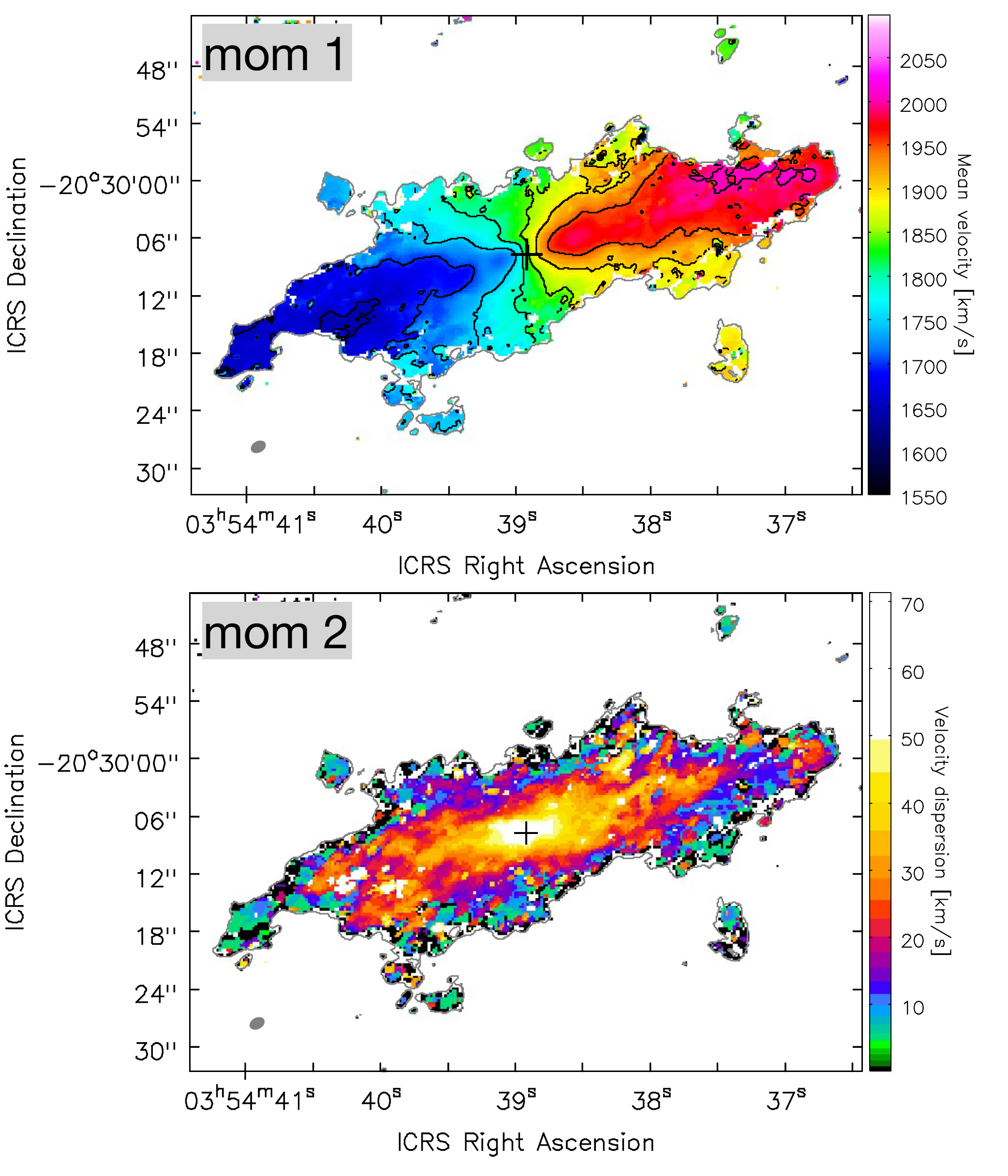}
 \caption{\emph{Top.} CO mean velocity \(\bar{v}\) (moment 1). The black contours are plotted from 1650 to 2000 km s\(^{-1}\) in steps of 50 km s\(^{-1}\). The ``+'' symbol marks the CO peak. \emph{Bottom.} CO velocity dispersion \(\sigma_v\) (moment 2).\label{fig:kin}}
\end{figure}

\subsection{Kinematics}\label{sec:kin}

Figure \ref{fig:kin} (top panel) shows the mean velocity (moment 1; velocity field) image of the galaxy derived from the CO data cube. The mean velocity is defined as the intensity-weighted velocity \(\bar{v}=\int{v\mathcal{S}_\mathrm{CO}\mathrm{d}v}/\int{\mathcal{S}_\mathrm{CO}\mathrm{d}v}\), and the computation was done over a range from 1558.62 to 2086.93 km s\(^{-1}\) (within \(\approx\pm260\) km s\(^{-1}\) relative to the systemic velocity) on a data cube that was clipped at \(3\sigma\). The observed velocity gradient shows that the molecular gas disk is rotating. The velocity field appears to be dominated by circular rotation, except in some regions related to the streamers (wind) in Figure \ref{fig:co}.

Figure \ref{fig:kin} (bottom panel) shows the intensity-weighted velocity dispersion (moment 2) defined as \(\sigma_v=\sqrt{\int\mathcal{S}_\mathrm{CO}(v-\bar{v})^2\mathrm{d}v/\int\mathcal{S}_\mathrm{CO}\mathrm{d}v}\). High velocity dispersion is observed toward the galactic center region, but also in some regions at the base of the streamers. The geometrical and kinematical properties of the disk and streamers are analyzed and modeled in detail in sections \ref{sec:gd} and \ref{sec:out}.

\subsection{Continuum Emission and Star Formation Rate}\label{sec:con}

\begin{figure*}
 \centering
  \includegraphics[width=1\textwidth]{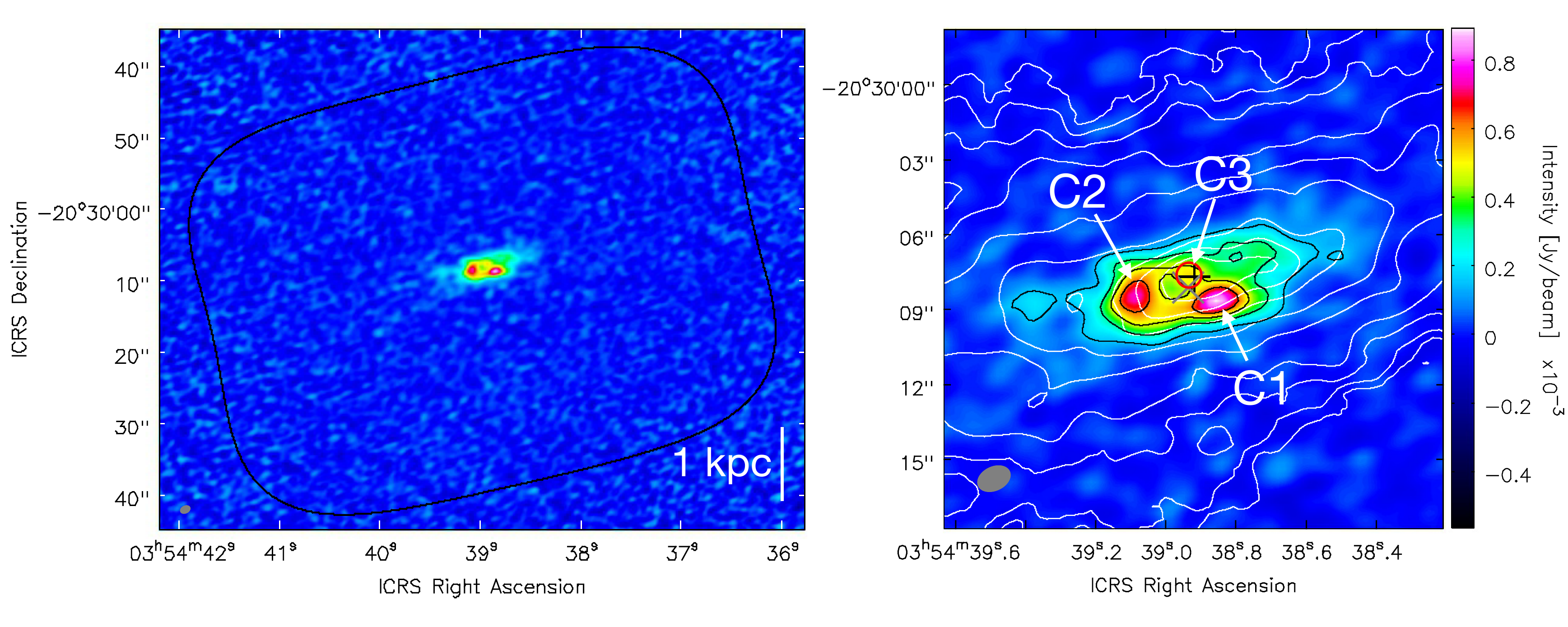}
 \caption{\emph{Left.} 100 GHz continuum intensity. \emph{Right.} Continuum intensity image and contours (black) plotted at \((5,10,15,20)\times1\sigma\), where \(1\sigma=31~\mu\mathrm{Jy~beam^{-1}}\). Also plotted are CO (1--0) contours (white) as in Figure \ref{fig:co}. The black ``+'' symbol marks the CO peak, the red circle marks the C3 source, and the gray ``x'' symbol marks the galactic center derived from a \(K_S\)-band image.\label{fig:con}}
\end{figure*}

Figure \ref{fig:con} shows the distribution of 100 GHz continuum in the entire field observed by ALMA (left panel) and in the central starburst region (right panel). The continuum was detected in the central 1 kpc (base of the superwind), where it is resolved into at least three major sources. The total flux density within a radius of \(r\leq10\arcsec\) (\(\approx950\) pc) is found to be \(S_\mathrm{100GHz}=11.3~\mathrm{mJy}\).

The coordinates of the continuum sources were determined by fitting over a region of diameter \(0\farcs949\) (beam minor axis) centered at the brightest pixel using \emph{imfit}. We identified three prominent sources, denoted by C1-C3, and their coordinates and peak intensities are given in Table \ref{tab:con}. The sources were identified as regions with highest peak flux densities, where C1 is the strongest, followed by C2 and C3.

The size of the continuum emission region is similar to those at 4860 and 8460 MHz in \citet{HS05} and \(11.3~\micron\) in \citet{Sie08}. The brightest compact source at low frequency is also spatially coincident with source C3 on our image. Interestingly, while C3 is the brightest source at 4860 and 8460 MHz, it is less prominent compared to C1 and C2 at 100 GHz. The spectral index of C3 is found to be \(\alpha\approx-1\) at low frequency, where \(S_\nu\propto\nu^\alpha\), suggesting non-thermal radiation. Since radiation in starbursts is typically dominated by synchrotron radiation at low frequency (\(\nu\lesssim10\) GHz) and free-free radiation at high frequency (\(\nu\sim100\) GHz) (e.g., \citealt{Con92}), the nature of these sources may be different. C1 and C2 may be massive \ion{H}{2} regions, whereas C3 is a nuclear starburst with synchrotron-radiating supernova remnants and/or an obscured low-luminosity AGN. However, there is no compelling evidence of an AGN from optical lines \citep{Kew00}; the nuclear activity appears to be dominated by stellar feedback. Note that C3 is spatially coincident with the CO peak and the brightest source at low radio frequency. As will be shown below, the distribution of predominantly old stars traced by the \(K_S\) band (gray ``x'' sign in the right panel of Figure \ref{fig:con}) is also consistent with this center.

In the starburst region, there are two hard (2-8 keV) X-ray sources detected by the \emph{Chandra X-ray Observatory} \citep{Str04a}. The coordinates of one of them, labelled ``X-ray west'' in \citet{HS05}, are coincident with our source C2. The other source has no prominent counterpart at 100 GHz.

Since continuum at 100 GHz is often dominated by thermal free-free emission in ionized gas, and ionized gas traces sites of recent star formation, such as \ion{H}{2} regions and supernova remnants, its flux density can be used to estimate the SFR in the nuclear starburst (e.g., \citealt{Ben15,Mic20}). Using the equations in these works, assuming an electron temperature of \(T_\mathrm{e}=5000~\mathrm{K}\) and that free-free radiation contributes 80\% to the flux at these frequencies, the measured continuum flux \(S_\mathrm{100GHz}\) yields a star formation rate of \(\mathrm{SFR_{ff}}\approx4~M_\sun~\mathrm{yr}^{-1}\) in the central 1 kpc. This value is consistent with the SFR derived from far-infrared luminosity, \(3.9~M_\sun~\mathrm{yr}^{-1}\) in \citet{Kew00}, and \(3.6~M_\sun~\mathrm{yr^{-1}}\) in \citet{Ken11}, who used H\(\alpha\) and \(24~\micron\) data.

\begin{table}
\begin{center}
\caption{Continuum Sources}\label{tab:con}
\begin{tabular}{lccc}
\tableline\tableline
Source & \(\alpha_\mathrm{ICRS}\) & \(\delta_\mathrm{ICRS}\) & Peak intensity \\
& & & (\(\mu\mathrm{Jy~beam^{-1}}\)) \\
\tableline
C1 & \(\mathrm{3^h54^m38\fs856}\) & \(\mathrm{-20\arcdeg30\arcmin08\farcs64}\) & \(894.2\pm5.4\) \\
C2 & \(\mathrm{3^h54^m39\fs086}\) & \(\mathrm{-20\arcdeg30\arcmin08\farcs48}\) & \(752.1\pm6.8\) \\
C3 & \(\mathrm{3^h54^m38\fs933}\) & \(\mathrm{-20\arcdeg30\arcmin07\farcs60}\) & \(517.6\pm6.8\) \\
\tableline
\end{tabular}
\end{center}
\end{table}

\section{Molecular Gas Disk}\label{sec:gd}

In section \ref{sec:mgd}, we stated that the CO streamers appear to be associated with the superwind. To study the molecular gas outflow, we first need to derive basic geometrical and kinematical parameters of the starburst disk, which is the base of the outflow. The parameters include the position angle, inclination, and rotational velocity. In this section, we make a model of a rotating disk, and then, based on the derived disk properties, separate the disk from the molecular gas entrained in the wind in section \ref{sec:out}.

\subsection{Disk Model}\label{sec:dgd}

Figure \ref{fig:co} (lower panel) shows that molecular gas is concentrated in a nearly edge-on disk that corresponds to the dust lane observed in optical images. We derived the basic geometric and kinematic parameters of the molecular gas disk using the code 3D-Based Analysis of Rotating Objects from Line Observations (\(^\mathrm{3D}\)Barolo; \citealt{DTF15}). The program fits a three-dimensional model to an emission-line data cube. The main advantage of the 3D approach over 2D methods is the fact that it can correct for beam smearing due to limited angular resolution even when the object is at high inclination. Computation in \(^\mathrm{3D}\)Barolo is done within elliptical rings in the cube, and the basic parameters include the galactic center coordinates, systemic velocity \(v_\mathrm{sys}\), inclination \(i\), and position angle PA. Setting the number of rings and \(\Delta R\), these parameters can be calculated for each ring in tilted-ring geometry (e.g., \citealt{RLW74}). Below, we describe the strategy of our analysis, starting from the parameters that were independently derived (position angle and galactic center) and then fixed during computation in \(^\mathrm{3D}\)Barolo.

\subsubsection{Position Angle}\label{sec:pa}

\begin{figure}
 \centering
  \includegraphics[width=0.45\textwidth]{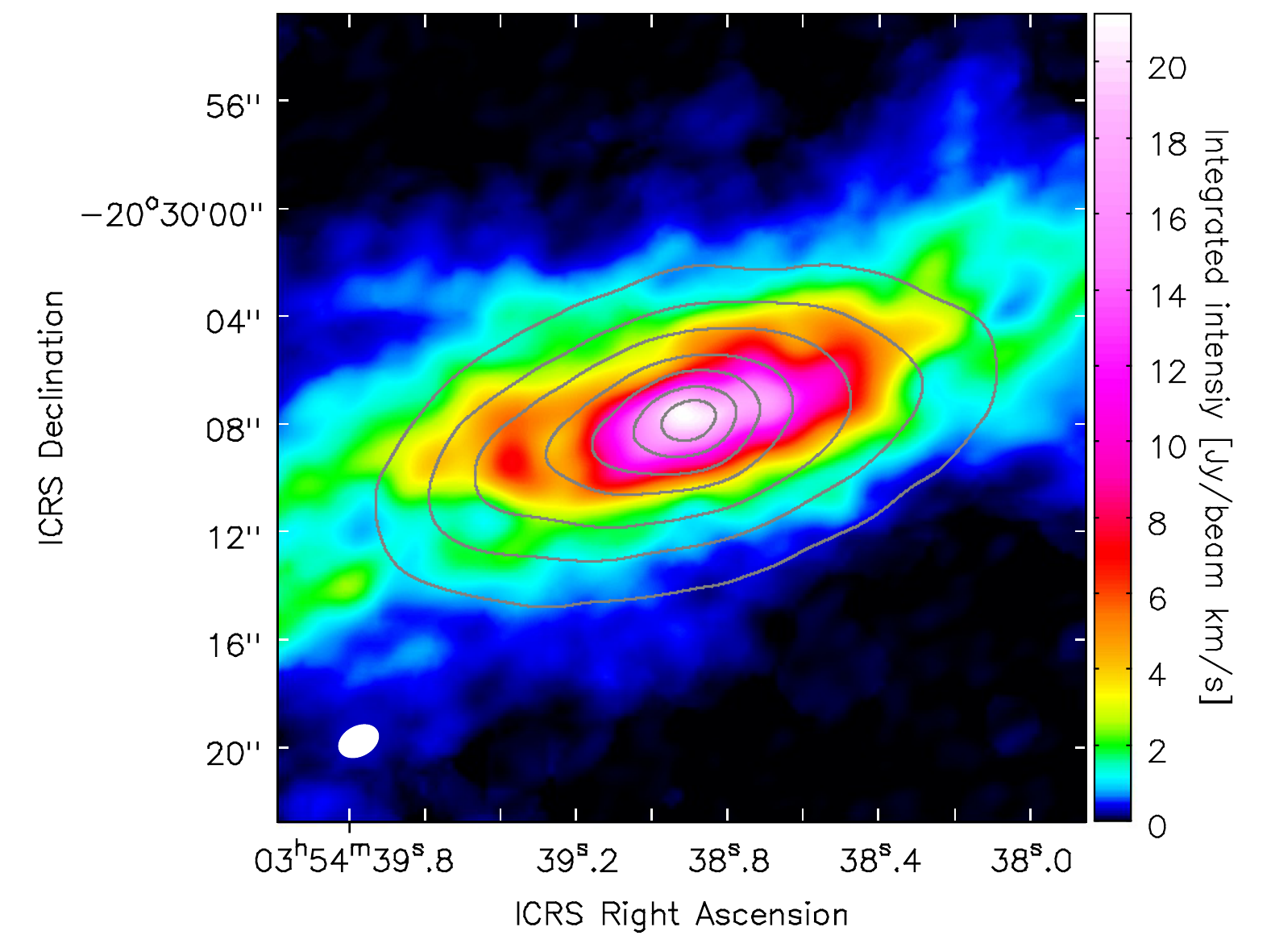}
 \caption{CO (1--0) integrated intensity image with \(K_S\)-band (2MASS) brightness contours in the central region. The contours are plotted at \((0.05,0.1,0.2,0.4,0.6,0.8,0.95)\times I_\mathrm{max}\) (maximum value).\label{fig:cok}}
\end{figure}

In order to limit the number of free parameters, we first estimated the position angle of the disk from a near-infrared image that traces old stellar populations (Figure \ref{fig:cok}). The position angle of the inner region of the galaxy was estimated from a Two Micron All Sky Survey (2MASS) \(K_S\)-band (wavelength \(2.00\mathrm{-}2.31~\micron\)) image \citep{Skr06}, which is not affected significantly by dust absorption. The image was fitted by an elliptical Gaussian function using the CASA task \emph{imfit} over a region of radius \(15\arcsec\) centered at the brightest pixel, yielding a position angle of \(\mathrm{PA}=-74\fdg35\pm0\fdg43\), in agreement with \(-74\fdg5\) in the 2MASS Large Galaxy Atlas \citep{Jar03}. In the following kinematical analysis, we adopt this value as the position angle of the galaxy and the molecular gas disk.

\subsubsection{Galactic Center}\label{sec:cen}

The coordinates of the \(K_S\)-band maximum obtained from the Gaussian fitting described above are \((\alpha,\delta)_\mathrm{ICRS}^{K_S}=(\mathrm{3^h54^m38\fs933,-20\arcdeg30\arcmin08\farcs08})\) (Table \ref{tab:gal}). This center, corresponding to the maximum concentration of old stars, is within \(0\farcs5\) from the CO and continuum (C3) peaks that were calculated in section \ref{sec:res} and given in Tables \ref{tab:gal} and \ref{tab:con}. Figure \ref{fig:cok} also shows that the maxima of CO and \(K_S\)-band images are spatially in good agreement. In the following computation, we adopt the CO peak as the center of the molecular gas disk, because it is spatially coincident though more prominent than the C3 continuum peak and has higher positional accuracy than the 2MASS center.

\begin{figure*}
 \centering
  \includegraphics[width=0.9\textwidth]{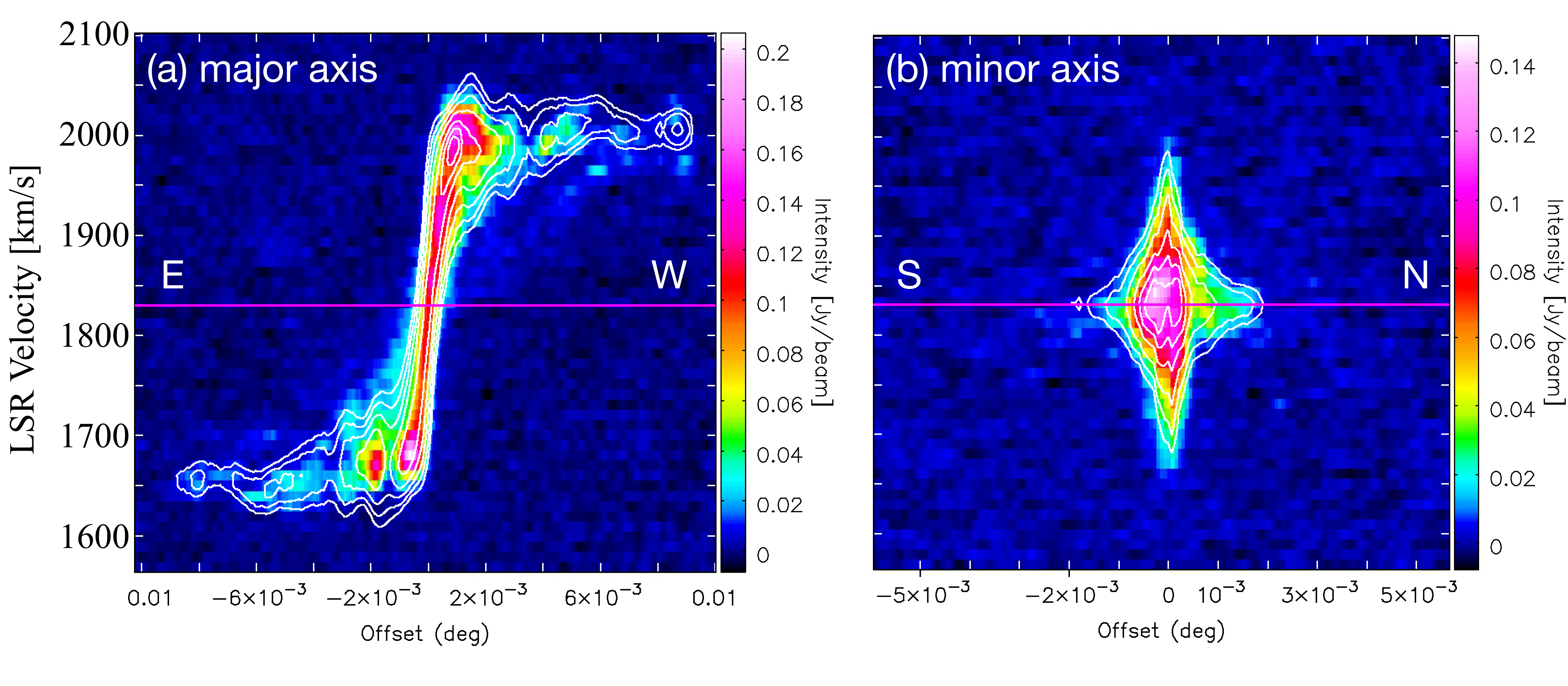}
 \caption{Position-velocity diagrams along the major (a) and minor (b) axes. The images are observational data and the contours are the \(^\mathrm{3D}\)Barolo model plotted at \((0.025,0.05,0.1,0.2,0.4,0.6,0.8)\times0.284~\mathrm{Jy~beam^{-1}}\) for the major axis and \((0.05,0.1,0.2,0.4,0.6,0.8)\times0.1795~\mathrm{Jy~beam^{-1}}\) for the minor axis. The horizontal line marks the systemic velocity \(v_\mathrm{sys}=1831~\mathrm{km~s^{-1}}\).\label{fig:pvds}}
\end{figure*}

\begin{figure*}
 \centering
  \includegraphics[width=0.9\textwidth]{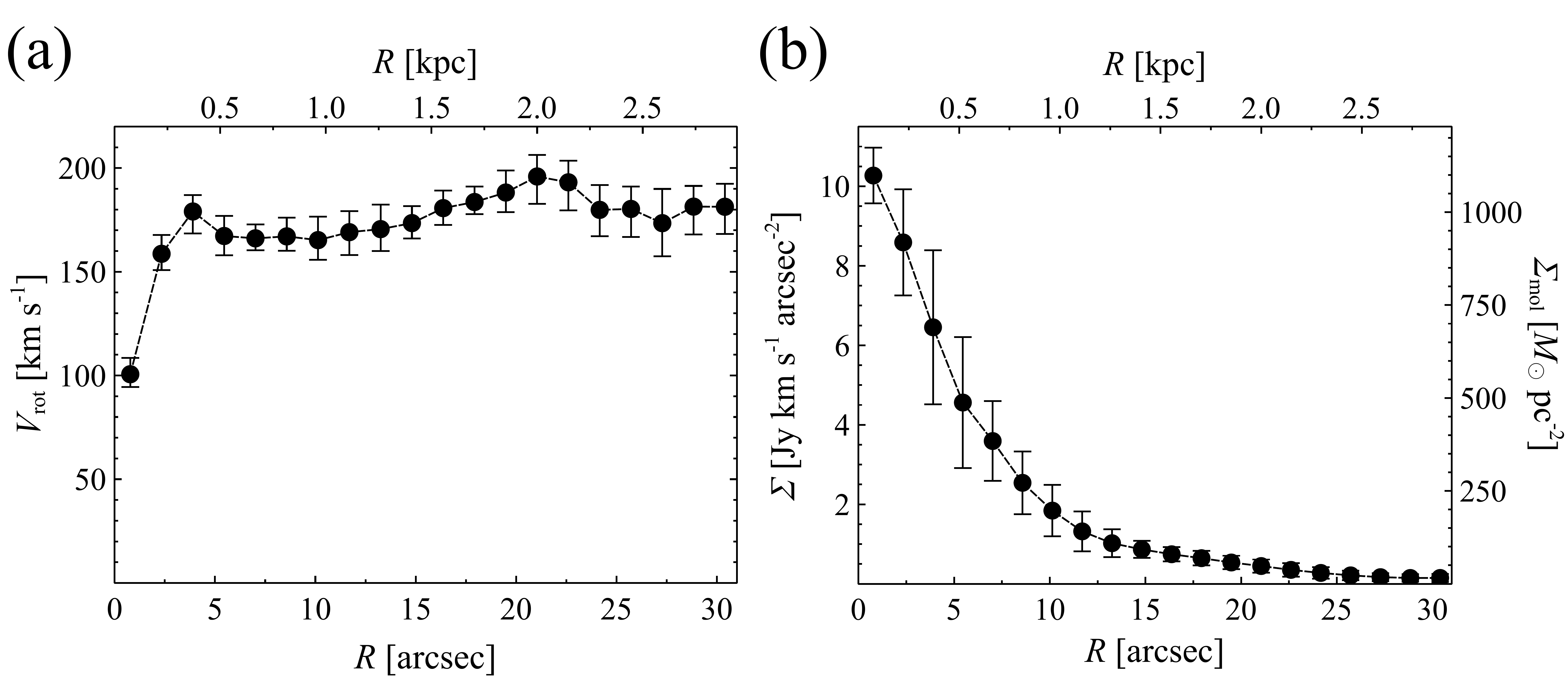}
 \caption{(a) Rotation curve \(v_\mathrm{rot}\). (b) Projected CO (1--0) surface brightness \(\Sigma\). The vertical axis on the right shows the mass surface density of molecular gas \(\Sigma_\mathrm{mol}\), which is corrected for the disk inclination.\label{fig:plo}}
\end{figure*}

\subsection{Two-Phase Computation and Results}\label{sec:com}

With position angle and galaxy center fixed, the computation was conducted in \(^\mathrm{3D}\)Barolo within 20 elliptical rings of radial width \(\Delta R=1\farcs56\) (beam major axis). The number of rings was chosen to cover the region with an outer radius of \(R\approx30\arcsec\) (\(\approx2.9\) kpc) within which CO was detected.

We first ran phase one calculations to estimate the disk inclination angle \(i\) and systemic velocity \(v_\mathrm{sys}\). We could not obtain \(i\) from other data, because the galaxy is early type and using the ratio of the minor to major galactic axis from an optical or near-infrared image would yield the axial ratio of the stellar bulge, which is not an appropriate value of \(i\) for a thin gaseous disk. Therefore, we used \(^\mathrm{3D}\)Barolo to determine the inclination for each ring, and the mean value was \(i=76\fdg1\) with a dispersion of \(\sigma=2\fdg4\). The initial-guess systemic velocity, derived from the global CO (1--0) spectrum as the midpoint between the velocities corresponding to the 20\% of the line maximum, was found to be \(v_\mathrm{sys}(\mathrm{LSR})=1831\pm5\) km s\(^{-1}\). The value is consistent with previous estimates: \(1850\pm20\) km s\(^{-1}\) (heliocentric) from \ion{H}{1} \citep{HS05}, \(1830\) km s\(^{-1}\) (LSR) \citep{Elf96} and \(1848\) km s\(^{-1}\) (heliocentric) \citep{You95} from single-dish CO observations, and \(1850\pm20\) km s\(^{-1}\) (heliocentric optical) from optical observations \citep{VR02}. The values of position angle, inclination, and systemic velocity were then fixed in phase two calculations that yielded the rotational velocity \(v_\mathrm{rot}\) and CO (1--0) surface brightness \(\Sigma\) for each ring. Note that our model is a set of coplanar rings with uniform circular motion in each ring.

The results of \(^\mathrm{3D}\)Barolo computations are presented in Figures \ref{fig:pvds} and \ref{fig:plo}. Figure \ref{fig:pvds} shows position-velocity (pv) diagrams (long-slit spectra) along the major and minor axes of the galaxy, where the position angle is the above value based on the \(K_S\)-band image. The image is the CO data cube and the contours are from the model cube created by the program. We can see that there are no large discrepancies between the data and the model in either direction, i.e., that the program successfully reproduced the data cube.

\subsection{Rotation Curve and Dynamical Mass}\label{sec:rcm}

The rotation curve of the disk, calculated by the program for rings separated by \(1\farcs56\), is shown in Figure \ref{fig:plo}(a). The rotational velocity increases steeply in the inner \(R\lesssim300\) pc (\(\approx3\arcsec\)) and then becomes flat. There are bumps at \(R\approx350\) pc and \(R\approx2\) kpc, but generally little variation in the curve to the outermost radius. A bump in the rotation curve may be an indicator of noncircular motions, e.g., arising from the presence of a bar or spiral arms. Under the approximation of spherically symmetrical mass distribution, the dynamical mass within the outermost radius (\(R\leq30\farcs401\) equivalent to \(\approx2.9\) kpc) is found to be \(M_\mathrm{dyn}=Rv_\mathrm{rot}^2/G=(2.21\pm0.32)\times10^{10}~M_\sun\), where the uncertainty includes the contribution from \(v_\mathrm{rot}\). The molecular gas mass therefore contributes 3--12\% of the dynamical mass in the central \(R\leq2.9\) kpc, depending on the adopted conversion factor \(X_\mathrm{CO}\) (section \ref{sec:mgd}).

Figure \ref{fig:plo}(b) shows the surface brightness of CO (1--0) emission \(\Sigma\) obtained in the elliptical rings. Note that the brightness steeply increases toward the galactic center; molecular gas has high central concentration. The vertical axis on the right is the corresponding mass surface density of molecular gas \(\Sigma_\mathrm{mol}\). The mass surface density was calculated by applying Equation \ref{eq:mol} to the surface brightness \(\Sigma\), using the standard Galactic conversion factor \(X_\mathrm{MW}=2.0\times10^{20}\mathrm{cm^{-2}~(K~km~s^{-1})^{-1}}\) and correcting for the inclination angle \(i\). The surface density peaks in the galactic center, with \(\Sigma_\mathrm{mol}=1100~M_\sun~\mathrm{pc^{-2}}\) within the innermost ring, and decreases with radius exponentially. The surface density is as low as \(\Sigma_\mathrm{mol}=15.9~M_\sun~\mathrm{pc^{-2}}\) in the ring of the outermost radius of 2.9 kpc.

\section{Molecular Gas Outflow}\label{sec:out}

In this section, we present evidence of a molecular gas outflow from the central 1 kpc starburst region. In sections \ref{sec:mor}-\ref{sec:dyn}, we analyze the morphology and dynamics of the outflow, and then in section \ref{sec:dis} discuss its energetics, relation with other phases of the ISM, and origin.

\subsection{Morphology}\label{sec:mor}

\begin{figure}
 \centering
  \includegraphics[width=0.45\textwidth]{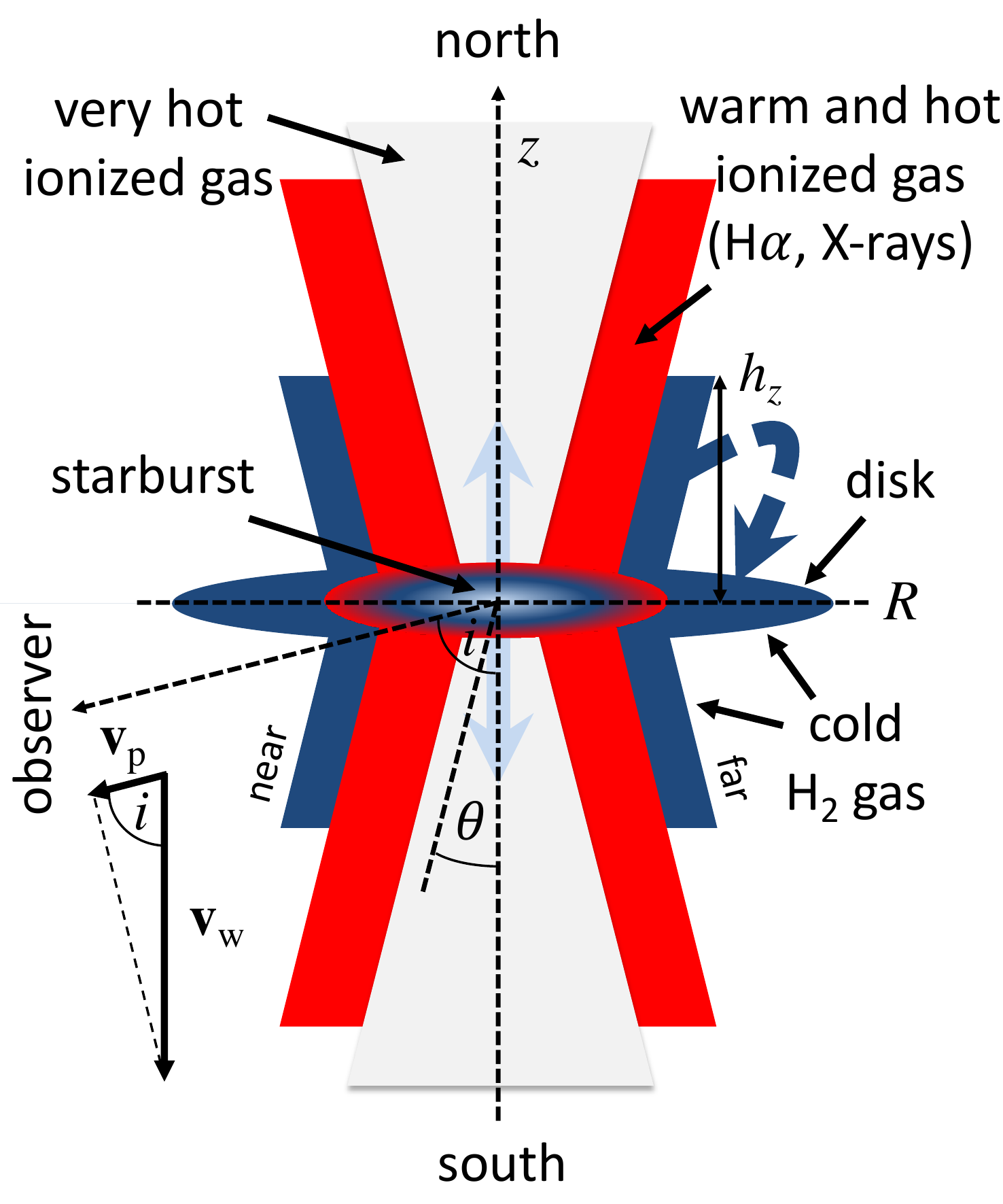}
 \caption{Cylindrically symmetrical multiphase gas outflow model (edge-on cross-section). The outflow velocity (parallel to \(z\)-axis), opening angle, and disk inclination are \(\textbf{v}_\mathrm{w}\), \(\theta\), and \(i\), respectively.}\label{fig:outm}
\end{figure}

In Figure \ref{fig:co}, we showed that the molecular gas exhibits streamers extending perpendicular to the galactic disk. \citet{HD99} found a similar structure in H\(\alpha\) emission and \citet{VR02} showed that warm ionized gas is flowing out of the central starburst region. The CO streamers suggest that molecular gas is also entrained in the wind, in which case the geometry of the outflow is biconical, the cone axis is close to or identical to the galactic minor axis (cylindrically symmetrical), and the two cones are truncated in the disk plane. The outflow is ``limb brightened'' and the observed streamers are the projected outflow walls. An illustration of such a model is shown in Figure \ref{fig:outm}, where the cold molecular (H\(_2\)) gas occupies an outer layer of a biconical wind that is cylindrically symmetrical about the \(z\)-axis (galactic minor axis). This is a modified version of the model first proposed by \citet{Nak87} to describe the CO outflow in M82; we also discussed it in \citet{Sal13} in the context of that galaxy. The model presented here distinguishes the hourglass-shaped distribution of warm ionized gas (H\(\alpha\) and soft X-rays) and the inner region filled with very hot (\(T\gtrsim10^7\) K), tenuous plasma, which is difficult to observe. In section \ref{sec:mul}, we will show that the structure of the outflow in NGC 1482 is indeed observed to be multilayered as illustrated in the figure.

In our analysis below, we will use geometrical parameters, including the vertical distance from the galactic disk \(h_z\), outflow opening angle \(\theta\), and disk inclination \(i\), as defined in Figure \ref{fig:outm}. The opening angle \(\theta\) is the angle between the \(z\)-axis and the velocity vector in the direction of the outflow. From the shape and orientation of the streamers in Figure \ref{fig:co}, the angle appears to be \(0\arcdeg\lesssim\theta\lesssim30\arcdeg\). The separation of the streamers (offset from \(z\) axis) indicates that the base of the outflow is wide, with a radius up to \(R\approx1.5\) kpc \textbf{(\(\approx16\arcsec\))}. The separation is larger than the size of the radio continuum region shown in Figure \ref{fig:con}.

\subsection{Kinematics}\label{sec:okin}

From the geometry of our outflow model, kinematical features of extraplanar gas flowing away from the galactic disk are expected to appear as a deviation of the observed mean velocity from the rotational velocity model produced in our \(^\mathrm{3D}\)Barolo modeling. Here, we investigate the distribution and magnitude of such deviations.

In cylindrical coordinates \((R,\phi,z)\), where the galaxy center is at \((R,z)=(0,0)\) and the galactic plane is defined by \(z=0\) (Figure \ref{fig:outm}), the observed velocity (Figure \ref{fig:kin}) of a disk inclined at an angle \(i\) is

\begin{equation}
\bar{v}=v_\mathrm{sys}+v_\phi\sin{i}\cos{\phi}+v_R\sin{i}\sin{\phi}+v_z\cos{i},
\end{equation}
where \(v_\mathrm{sys}\) is the systemic velocity, \(v_\phi\) is the azimuthal velocity, \(v_R\) is the radial velocity, and \(v_z\) is the vertical velocity in the direction perpendicular to the galactic disk. In general, all three velocity components are functions of \((R,\phi,z)\). A model rotational velocity, where only coplanar (\(z=0\)), uniform circular motion is included, can be expressed as

\begin{equation}\label{eq:mod}
v_\mathrm{mod}=v_\mathrm{sys}+v_\mathrm{rot}\sin{i}\cos{\phi},
\end{equation}
where \(v_\mathrm{rot}(R)\) is the rotational velocity that does not depend on \(\phi\), such as the one plotted in Figure \ref{fig:plo}(a). The residual velocity, \(v_\mathrm{res}=\bar{v}-v_\mathrm{mod}\), is then

\begin{equation}\label{eq:res}
v_\mathrm{res}=(v_\phi-v_\mathrm{rot})\sin{i}\cos{\phi}+v_R\sin{i}\sin{\phi}+v_z\cos{i}.
\end{equation}
In the absence of radial motions, \(v_R=0\) at all radii, so

\begin{equation}
v_\mathrm{res}=(v_\phi-v_\mathrm{rot})\sin{i}\cos{\phi}+v_z\cos{i}.
\end{equation}
If the model reproduces the rotation of the galaxy, the term \(v_\phi-v_\mathrm{rot}\approx0\), and we are left with the \(z\)-component velocity \(v_\mathrm{res}\approx v_z\cos{i}\).

\begin{figure}
 \centering
  \includegraphics[width=0.5\textwidth]{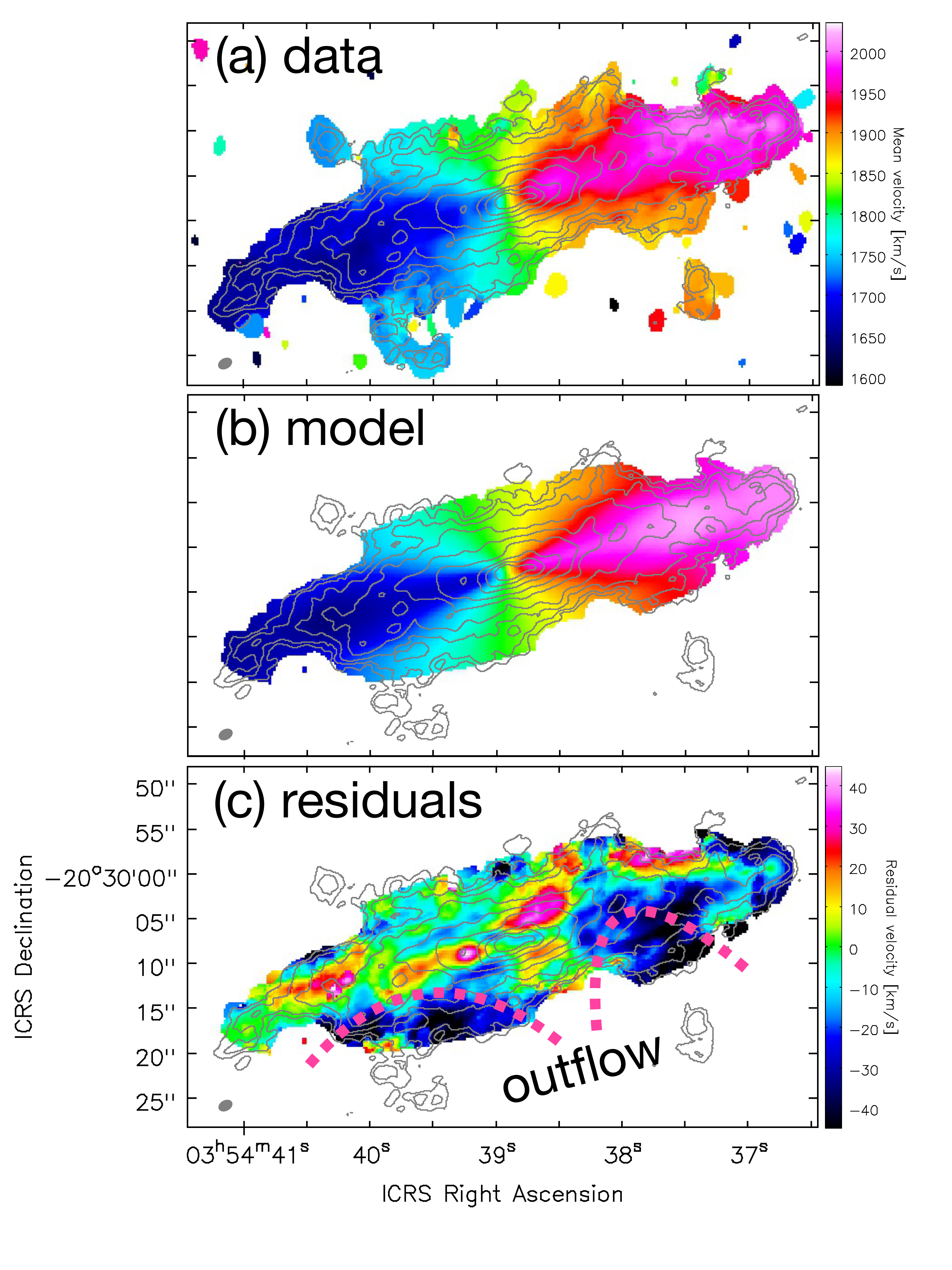}
 \caption{(a) Mean velocity \(\bar{v}\) (moment 1) image. (b) Model velocity \(v_\mathrm{mod}\). (c) Residual velocity \(\bar{v}-v_\mathrm{mod}\). The contours are as in Figure \ref{fig:co}. Dotted magenta lines indicate the regions dominated by velocities blueshifted with respect to the rotating disk as expected for the approaching side of an outflow from an inclined disk.\label{fig:res}}
\end{figure}

Figure \ref{fig:res} shows: (a) the observed mean velocity (moment 1 image) \(\bar{v}\), (b) model rotational velocity \(v_\mathrm{mod}\) (Equation \ref{eq:mod}), as a two-dimensional representation of the velocity in Figure \ref{fig:plo}(a), and (c) the residual velocity image calculated from \(v_\mathrm{res}=\bar{v}-v_\mathrm{mod}\). We find that most of the residuals throughout the disk are within \(|v_\mathrm{res}|\leq20~\mathrm{km~s^{-1}}\), which can be explained by noncircular motions in the disk (\(v_R\neq0\)). Along the major axis at \(z=0\), we note in the residual velocity image in Figure \ref{fig:res}(c) that there are redshifted and blueshifted regions on opposite sides of the galactic center at radius \(R\approx10\arcsec\) (1 kpc). This pattern is consistent with inflow motion and indicates that within the disk there is a mechanism for efficient fueling of the central starburst. Other notable exceptions from circular motion are the blueshifted regions (\(v_\mathrm{res}<0\)) southeast and southwest of the center, marked by magenta dotted lines. Considering the orientation of the galactic disk (Figure \ref{fig:outm}), these blueshifted features can be explained as an outflow toward the observer (\(v_z<0\)). If the outflow is perpendicular to the galactic plane, the observed velocity will be blueshifted in the south with respect to the velocity in the rotating disk, resulting in a mean velocity that deviates from the model \(v_\mathrm{mod}\). Maximum observed residuals are \(|v_\mathrm{res}|\approx40~\mathrm{km~s^{-1}}\), but typical values are \(20\lesssim |v_\mathrm{res}|\lesssim 30\) km s\(^{-1}\). From the above equation, the vertical velocity component is then \(80\lesssim |v_z|\lesssim120\) km s\(^{-1}\). The blueshifted components are at a radius up to \(R\lesssim2\) kpc (\(\approx21\arcsec\)) from the center and located at the basis of the wind streamers indicated by arrows in Figure \ref{fig:co} (lower panel). On the northern side, the residual velocities are smaller, but there are some places especially northwest of the center with redshifted velocities up to \(v_\mathrm{res}\approx+30~\mathrm{km~s^{-1}}\), as expected from the receding side of the outflow. Note that the NW side of the outflow in Figure \ref{fig:res}(c) also contains a blueshifted feature that appears to emerge from the disk and is surrounded by redshifted gas. This streamer may have been ejected from the disk at a high opening angle \(\theta\) so that it appears blueshifted with respect to the generally redshifted outflow gas.

\begin{figure}
 \centering
  \includegraphics[width=0.45\textwidth]{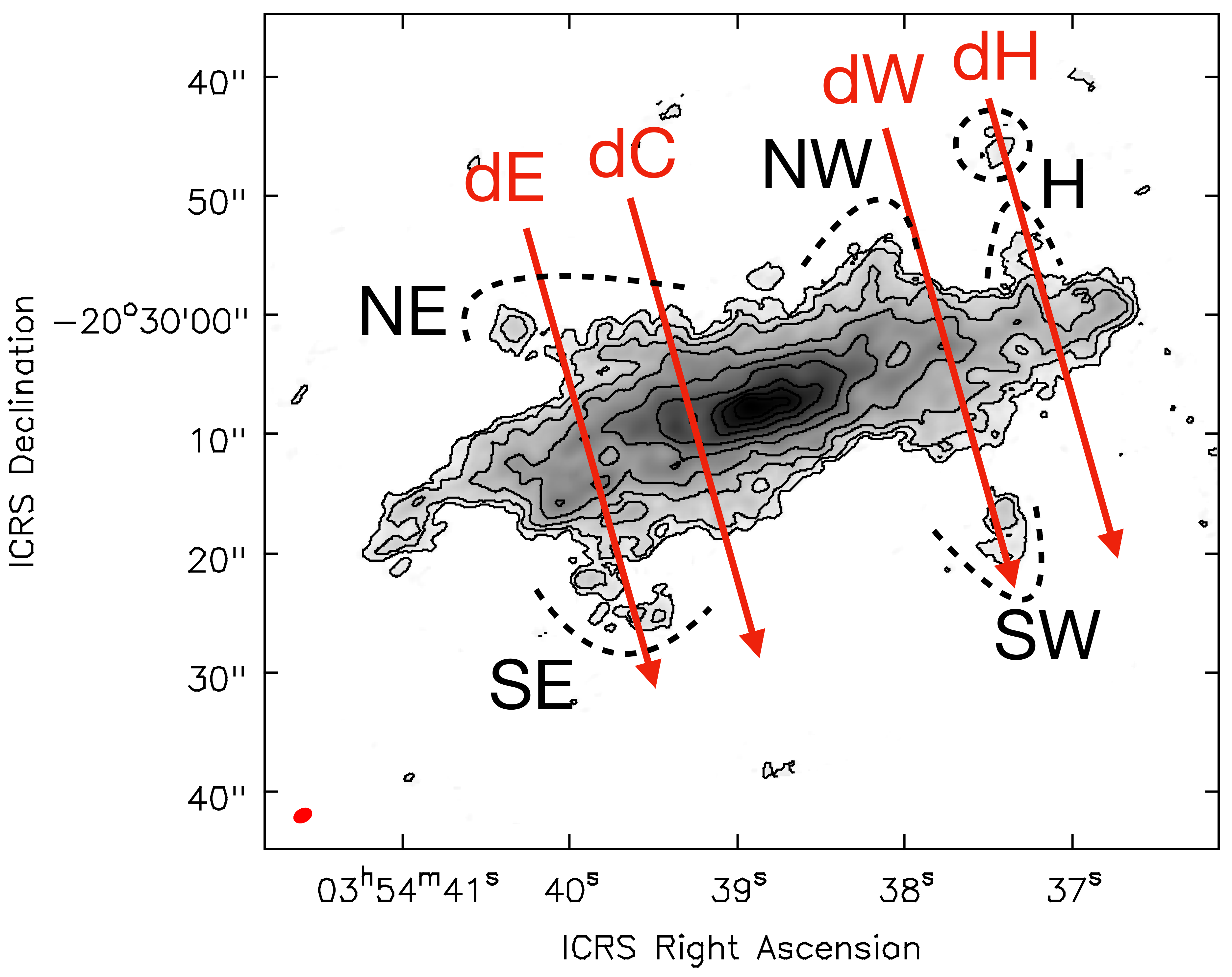}
 \caption{CO (1--0) integrated intensity map with pv diagram directions (arrows): dE (\(\Delta x=-14\arcsec\)), dC (\(\Delta x=-5\arcsec\)), dW (\(\Delta x=17\arcsec\)), and dH (\(\Delta x=26\arcsec\)), where \(\Delta x\) is the offset from center along the major axis (\(x>0\) toward west). The cuts are \(40\arcsec\) long, parallel to the minor axis, and their widths are 25, 5, 25, and 15 pixels, respectively, where the pixel size is \(0\farcs25\). Streamers are indicated by dashed lines. Also shown are features H.}\label{fig:pvd1}
\end{figure}

\begin{figure*}
 \centering
  \includegraphics[width=1\textwidth]{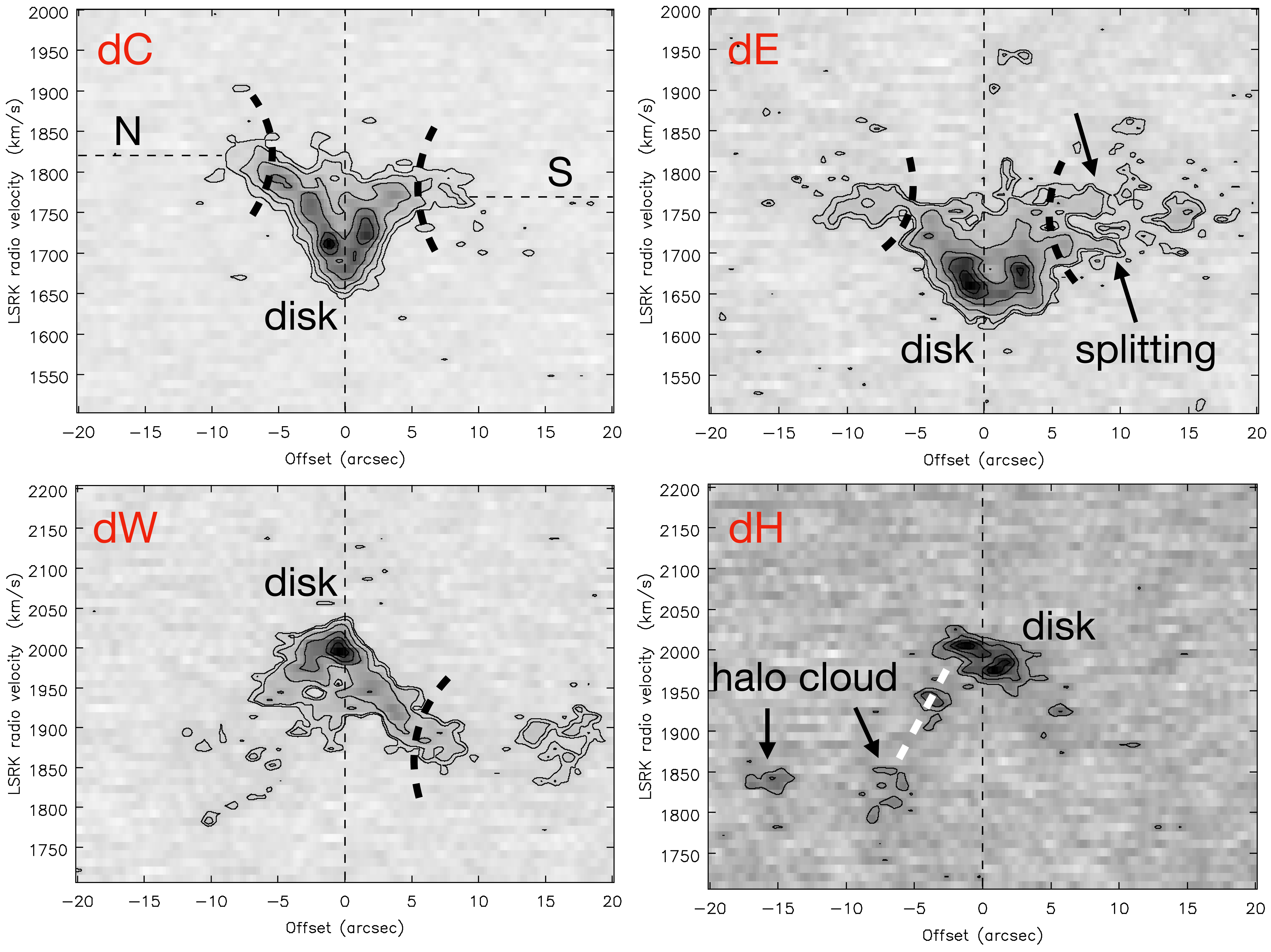}
 \caption{Position-velocity diagrams of Figure \ref{fig:pvd1}. Bold black dashed lines mark the approximate boundaries between the disk and outflow regions. Note the velocity difference between N and S sides in panel dC. Panel dE reveals line splitting with a separation of 60 km s\(^{-1}\). The white dashed line in panel dH indicates a steep velocity gradient in the region where the halo gas is approaching the disk. The contours are plotted at \((0.05,0.1,0.2,0.4,0.6,0.8)\times89.4~\mathrm{mJy~beam^{-1}}\) (dC), \((0.07,0.1,0.2,0.4,0.6,0.8)\times32.7~\mathrm{mJy~beam^{-1}}\) (dE), \((0.07,0.1,0.2,0.4,0.6,0.8)\times36.2~\mathrm{mJy~beam^{-1}}\) (dW), and \((0.2,0.4,0.6,0.8)\times18.2~\mathrm{mJy~beam^{-1}}\) (dH).}\label{fig:pvd2}
\end{figure*}

To estimate the outflow velocity \(v_\mathrm{w}\), we also investigate the kinematics of molecular gas in position-velocity space. Position-velocity (pv) diagrams obtained from cuts perpendicular to the galactic disk are shown in Figures \ref{fig:pvd1} and \ref{fig:pvd2}. In Figure \ref{fig:pvd1}, the outflow streamers that correspond to the above blueshifted components are denoted by SE and SW and the corresponding diagrams by dE and dW. The pv diagram dC is offset from the galactic center by \(\Delta x=-5\arcsec\) in order to include the extended emission detected parallel to the minor axis and which is less prominent in a pv diagram at \(\Delta x=0\arcsec\) through the galactic center (Figure \ref{fig:pvds}).

Figure \ref{fig:pvd2} (panel dC) shows velocity components in the south (S) and north (N), and the projected difference between the two is found to be \(|\Delta v_\mathrm{p}|\approx50~\mathrm{km~s^{-1}}\). In the case that the outflow opening angle is \(\theta>0\arcdeg\), we would expect to observe line splitting into two projected velocity components at \(v_\mathrm{p,near}=v_\mathrm{w}\cos(i-\theta)\) and \(v_\mathrm{p,far}=v_\mathrm{w}\cos(i+\theta)\) relative to the systemic velocity, where ``near'' and ``far'' are as shown in Figure \ref{fig:outm}, and \(|v_\mathrm{p,near}-v_\mathrm{p,far}|=2v_\mathrm{w}\sin{i}\sin{\theta}\) would be the observed velocity difference between the two components. Since we do not observe line splitting toward N and S in panel dC, the opening angle may be \(\theta\approx0\arcdeg\) in this region. For simplicity, we assume that the outflow observed in panel dC is perpendicular to the galactic disk and derive the deprojected outflow velocity from \(v_\mathrm{w}=|\Delta v_\mathrm{p}|/(2\cos{i})\), where \(|\Delta v_\mathrm{p}|=2v_\mathrm{p}\) (see Figure \ref{fig:outm}). For the projected N-S velocity difference of \(|\Delta v_\mathrm{p}|\approx50~\mathrm{km~s^{-1}}\) in dC and the molecular disk inclination of \(i=76\arcdeg\) derived using \(^\mathrm{3D}\)Barolo, we obtain \(v_\mathrm{w}\approx100~\mathrm{km~s^{-1}}\). This is consistent with \(v_z\), calculated above from the average residual velocity in the streamers and we adopt it as the representative velocity of the molecular gas outflow. From the overall morphology of the streamers, the opening angle of the outflow may be somewhat larger at some places, of the order of \(\approx10\arcdeg\mathrm{-}20\arcdeg\). The pv diagrams also indicate that the outflow velocity is approximately constant in the inner \(|z|<1.5\) kpc \textbf{(\(\approx16\arcsec\))}.

Panels dE and dW in Figure \ref{fig:pvd2} show pv diagrams parallel to the minor galactic axis and running through the SE and SW streamer regions. Here too, we generally find that the projected velocity is lower in the south than in the north, as expected from an outflow. Note that line splitting is observed in streamer SE (panel dE), with a separation of \(\Delta v\approx60~\mathrm{km~s^{-1}}\). It is possible that the two features arise from streamers ejected at different opening angles, one in front of another, with different projected velocities. On the other hand, the feature could also indicate an expanding superbubble. Superbubbles observed as CO-bright shells in outflows are also confirmed in other superwind galaxies, such as M82 \citep{Wei99,Mat00} and NGC 253 \citep{Bol13}, and appear to be a common feature of starburst-driven outflows (e.g., \citealt{MM88,Kim17}).

Panel dH in Figure \ref{fig:pvd2} reveals a feature that does not appear to be related to any of the outflow streamers. The velocities of the halo clouds denoted by H in Figure \ref{fig:pvd1} are different from what would be expected from gas outflow observed north of the galactic disk that should be receding from us. The velocities are close to the systemic velocity \(v_\mathrm{sys}\) suggesting that this extraplanar gas may be moving in the plane of the sky and possibly toward the disk. In other words, these features, and especially the one marked by a dotted line close to the disk, can be interpreted as gas inflow from the halo. The inflow scenario can be understood from geometry and velocity: for an outflow from an inclined disk as shown in Figure \ref{fig:outm}, gas should be receding relative to an observer in the galaxy rest frame on the N side and approaching on the S side. However, we see gas approaching on the N side too. Figure \ref{fig:pvd2} (panel dH) reveals a discontinuous bridge (white dotted line) between the disk component (denoted by ``disk'') at \(v_\mathrm{LSR}\approx2000~\mathrm{km~s^{-1}}\) and a halo cloud at an offset of \(-7\arcsec\) and velocity \(v_\mathrm{LSR}\approx1830~\mathrm{km~s^{-1}}\). The velocity difference between the disk and halo gas is \(170~\mathrm{km~s^{-1}}\), and since there is no evidence for an energy source (e.g., large \ion{H}{2} region) in this part of the disk that could eject the gas at such speed, the halo cloud is likely undergoing a collision with the disk -- a possible evidence of a ``fountain flow'' created by the outflow (e.g., \citealt{KO18}). An illustration of the trajectory of a halo cloud raining down onto the disk is shown in Figure \ref{fig:outm} as a an arrow that points toward the disk. In section \ref{sec:evi} below we calculate the mass inflow rate based on the kinematics and mass of these extraplanar clouds.

Another possibility is that these halo clouds are tidal debris inflowing from the direction of the nearby galaxy NGC 1481, which is located at a projected distance of \(\approx300\arcsec\) (\(\approx28.5\) kpc) toward northwest. There is some evidence of tidal features in the outer disk of NGC 1482 (\(\approx7\) kpc northwest of the center), possibly a consequence of tidal interaction \citep{Kim12}. Such a gravitational disturbance in the past may also explain the presence of a large amount of molecular gas in the central region of NGC 1482 that fuels the starburst. We will discuss this feature further in section \ref{sec:ori}.

\subsection{Dynamics}\label{sec:dyn}

Next, we calculate the kinetic energy, mass outflow rate, and mass-loading factor of the molecular outflow. The latter is defined as the ratio of the mass outflow rate to the star formation rate, \(\dot{M}_\mathrm{w}/\mathrm{SFR}\), and can serve as an indicator of negative feedback on star formation (e.g., \citealt{Chi17,RB20}). From the CO distribution in the streamers (Figure \ref{fig:pvd1}), disk geometry, and the kinematical model presented above, we estimate that the CO (1--0) flux in the outflow is \(\int S_\mathrm{CO}\mathrm{d}v\approx70~\mathrm{Jy~km~s^{-1}}\) with an uncertainty of \(\approx30\%\). This is approximately 10\% of the total CO (1--0) flux in NGC 1482. Applying a low CO-to-H\(_2\) conversion factor of \(X_\mathrm{CO}=0.5\times10^{20}\mathrm{cm^{-2}(K~km~s^{-1})^{-1}}\), equivalent to \(\alpha_\mathrm{CO}=1~M_\sun(\mathrm{K~km~s^{-1}~pc^2})^{-1}\), as a representative of an outflow gas with moderate optical depth (e.g., \citealt{BWL13,Zsc18}), the total molecular gas mass in the outflow is \(M_\mathrm{w}\sim7\times10^7~M_\sun\). Adopting a deprojected outflow velocity of \(v_\mathrm{w}=100~\mathrm{km~s^{-1}}\), the kinetic energy of the molecular wind becomes \(E_\mathrm{w}=M_\mathrm{w}v_\mathrm{w}^2/2\sim7\times10^{54}~\mathrm{erg}\). By comparison, the mass and kinetic energy of the warm ionized gas (\ion{H}{2}) outflow traced by H\(\alpha\) are \(M_\mathrm{w,H\alpha}\gtrsim3.6\times10^5~n^{-1}_\mathrm{e,2}~M_\sun\) and \(E_\mathrm{w,H\alpha}\gtrsim2\times10^{53}~n^{-1}_\mathrm{e,2}\) erg, respectively, where \(n_\mathrm{e,2}\) is the electron density normalized as \(n_\mathrm{e,2}=n_\mathrm{e}/100\) cm\(^{-3}\) \citep{VR02}. The ionized gas outflow is defined in that work as the region where the intensity ratio \([\text{\ion{N}{2}}]\lambda 6585/\mathrm{H}\alpha>1\). For an electron density of \(n_\mathrm{e}\sim10\) cm\(^{-3}\), the ionized gas outflow mass is an order of magnitude lower than that of molecular gas. The [\ion{S}{2}]\(\lambda\)6733/\(\lambda\)6718 doublet ratio map in \citet{SB10} suggests that the electron density of \(n_\mathrm{e}\sim10\) cm\(^{-3}\) is representative for the warm ionized gas in the wind within the inner 1 kpc, but this value is uncertain.

The mass outflow rate from the starburst disk can be obtained from the continuity equation, \(-\partial \rho/\partial t=\nabla\cdot(\rho\textbf{v}_\mathrm{w})\), where \(\rho\) is the gas density and \(\textbf{v}_\mathrm{w}\) is the outflow velocity vector. Using the divergence theorem and defining \(\dot{M}_\mathrm{w}\equiv -\partial M/\partial t\) as the mass outflow rate, the continuity equation can be expressed as

\begin{equation}\label{eq:ce}
\dot{M}_\mathrm{w}=\int_A\rho\mathbf{v}_\mathrm{w}\cdot\mathrm{d}\mathbf{A},
\end{equation}
where \(A\) is the area through which gas is flowing. The above equation tells us that the mass lost from the central disk is equal to the mass filling the outflow. For a cylindrically shaped outflow with a constant velocity parallel to the \(z\)-axis (\(v_\mathrm{w}=v_z=\mathrm{constant}\); in section \ref{sec:okin} it was found that the opening angle is \(\theta\approx0\arcdeg\), which we adopt here as well) and uniform density, the mass outflow rate from Equation \ref{eq:ce} is

\begin{equation}\label{eq:mor}
\dot{M}_\mathrm{w}=v_\mathrm{w}\frac{M_\mathrm{w}}{h_z},
\end{equation}
where \(M_\mathrm{w}\) is the total outflow mass within \(|z|\leq h_z\). For \(h_z=1\) kpc, the distance that encompasses most of the outflow gas, Equation \ref{eq:mor} yields \(\dot{M}_\mathrm{w}\sim7~M_\sun~\mathrm{yr}^{-1}\).

Note that for a spherically symmetrical, uniformly filled outflow with \(v_\mathrm{w}=v_r\), the mass outflow rate from Equation \ref{eq:ce} becomes \(\dot{M}_\mathrm{w}=3v_\mathrm{w}M_\mathrm{w}/h_r\), where \(h_r\) is the outflow radius. The morphology and kinematics in NGC 1482 suggest that the gas outflow is approximately cylindrical, so we adopt the cylinder geometry that gives a conservative estimate of \(\dot{M}_\mathrm{w}\).

Under the approximation that the outflow velocity is constant, the dynamical timescale is \(t_\mathrm{w}=h_z/v_\mathrm{w}\sim1\times10^7~\mathrm{yr}\). By comparison, \citet{VR02} obtained a dynamical timescale of \(6\times10^6~\mathrm{yr}\) for the warm ionized gas, that moves at a higher velocity of \(v_\mathrm{w,H\alpha}\approx250~\mathrm{km~s^{-1}}\). Their results are consistent with our derived timescale.

The value of \(\dot{M}_\mathrm{w}\) is of the same order of magnitude as the SFR in the central 1 kpc derived in section \ref{sec:con} (\(\mathrm{SFR}\approx4~M_\sun~\mathrm{yr}^{-1}\)) and larger than the mass outflow rate of the warm ionized gas estimated from optical lines, which is \(\dot{M}_\mathrm{w,H\alpha}\sim0.6~M_\sun~\mathrm{yr^{-1}}\) for an electron density of \(n_\mathrm{e}\sim10\) cm\(^{-3}\). Thus, the molecular outflow has a mass-loading factor of \(\dot{M}_\mathrm{w}/\mathrm{SFR}\sim2\).

\section{Discussion}\label{sec:dis}

\subsection{Starburst-Driven Outflow}\label{sec:sdo}

What is driving the superwind in NGC 1482? The large spatial extent of the outflow base (1 kpc), vigorous star formation, and the absence of an energetically relevant AGN suggest that the outflow is driven by starburst activity. In that case, the input energy is provided by two main sources (e.g., \citealt{Lei99,NS09}): core-collapse (Type II) supernova explosions \citep{CC85} that eject material (and also accelerate cosmic rays), and massive O and B stars in young star clusters that produce radiation pressure including stellar winds \citep{MQT05,MMT11}. In fact, massive stars first produce radiation pressure and only after a few million years explode as supernovae. For a starburst episode that produces supernova explosions during time \(t_\mathrm{SN}\), the kinetic energy transferred to the ISM from the explosions can be expressed as

\begin{equation}
E_\mathrm{SN}=E_0\int_{-t_\mathrm{SN}}^0\mathcal{R}_\mathrm{SN}(t)\xi(t)\mathrm{d}t,
\end{equation}
where \(E_0=10^{51}~\mathrm{erg}\) is the total energy output from a single core-collapse supernova explosion, \(\mathcal{R}_\mathrm{SN}\) is the supernova rate that depends on star formation history, \(\xi\) is the efficiency of energy transfer to the ISM, and integration is carried out from the time the supernovae started exploding, \(t=-t_\mathrm{SN}\), to the present, \(t=0\). The fraction of energy (\(1-\xi\)) which is not transferred to the ISM is assumed to dissipate mainly by radiation, but the value of \(\xi\) is generally difficult to estimate (e.g., \citealt{SH09,Sim15,WN15}). The core-collapse supernova rate in NGC 1482 is estimated to be \(\mathcal{R}_\mathrm{SN}=0.2L_\mathrm{IR}/10^{11}(L_\sun)=0.10~\mathrm{yr}^{-1}\) \citep{Str04b} from IR luminosity and 0.14 yr\(^{-1}\) from non-thermal radio continuum \citep{HS05}. \citet{Kew00} derived a lower value of 0.03 yr\(^{-1}\). For a timescale of \(t_\mathrm{SN}\approx t_\mathrm{w}=1\times10^7~\mathrm{yr}\) and assumed constant \(\mathcal{R}_\mathrm{SN}=0.03\mathrm{-}0.14~\mathrm{yr}^{-1}\) and \(\xi\sim0.1\) \citep{MQT05}, the kinetic energy transferred to the ISM in the starburst region is \(E_\mathrm{SN}\approx0.3\mathrm{-}1.4\times10^{56}~\mathrm{erg}\). The mechanical energy injection to the ISM is thus \(\sim10\) times larger than the kinetic energy carried by the molecular outflow. By comparison, the kinetic energy of the warm ionized gas outflow  traced by H\(\alpha\) is estimated to be \(E_\mathrm{w,H\alpha}\gtrsim2\times10^{53}n^{-1}_\mathrm{e,2}~\mathrm{erg}\) \citep{VR02}, which is \(\gtrsim1\%\) of \(E_\mathrm{SN}\), similar to the values of 1--20\% that have been observed in outflows of warm ionized gas in a number of other nearby galaxies (e.g., \citealt{Chi17}). Thus, the total kinetic energy of the molecular and warm ionized gas outflows in NGC 1482 is \(\sim1\%\) of the total energy released by supernova explosions and \(\sim10\%\) of \(E_\mathrm{SN}\), the energy transferred to the ISM for an efficiency of \(\xi=0.1\) (10\%). A large fraction of the remaining energy may be in other wind phases, which are difficult to measure, such as the tenuous gas at \(T\gtrsim10^7\) K. Since molecular gas is one of the dominant phases of the outflow energy budget, we conclude that the energy released by starburst activity is sufficient to drive the superwind in NGC 1482.

It was mentioned above that the efficiency \(\xi\) is largely uncertain. While some of the supernova energy is radiated away in the cooling process, the momentum in the starburst and wind must be conserved. The total momentum deposition of supernova ejecta is

\begin{equation}
p_\mathrm{SN}=m_0v_0\int_{-t_\mathrm{SN}}^0\mathcal{R}_\mathrm{SN}(t)\mathrm{d}t,
\end{equation}
where \(m_0\approx10~M_\sun\) is the typical ejecta mass, and \(v_0\approx3000\) km s\(^{-1}\) its velocity. These numbers refer to initial conditions. For \(t_\mathrm{SN}=1\times10^7\) yr\(^{-1}\) and \(\mathcal{R}_\mathrm{SN}=0.1\) yr\(^{-1}\), we get \(p_\mathrm{SN}\approx3\times10^{10}~M_\sun~\mathrm{km~s^{-1}}\). The final momentum may actually be larger, \(\approx3\times10^5~M_\sun~\mathrm{km~s^{-1}}~n_0^{-0.17}\) per supernova, where \(n_0\) is the ambient gas density, as the ejecta evolve while propagating through the ISM \citep{KO15,MFQ15}. On the other hand, a fraction of momentum will be cancelled at the interface of shell-shell collisions of different supernova ejecta in the starburst disk and not contribute to vertical flow. The overall molecular wind momentum is \(p_\mathrm{w}= M_\mathrm{w}v_\mathrm{w}\sim7\times10^9~M_\sun~\mathrm{km~s^{-1}}\), which is \(\sim20\%\) of the initial momentum deposited by supernovae. The rest of the momentum may be distributed among the other ISM phases in the wind and starburst disk.

Given the large uncertainties involved, the energetics is comparable to that found in some other nearby starburst-driven winds (e.g., \citealt{Tsa12,PS16}).

\subsection{Is NGC 1482 Losing Mass?}\label{sec:evi}

Galactic outflows are essential to transporting metal-enriched interstellar gas to the CGM and IGM (e.g., \citealt{SH09,Rup19}). Is the wind in NGC 1482 capable of ejecting a significant fraction of the ISM and thereby quenching star formation? Below, we carry out a simple analysis of a ballistic outflow with gravity as the dominant force acting on cold gas.

Assuming that the mass distribution is spherically symmetrical, the escape velocity at a distance \(r\) from the galactic center is

\begin{equation}\label{eq:esc}
v_\mathrm{esc}(r)=\sqrt{2|\Phi(r)|},
\end{equation}
where \(\Phi(r)\) is the gravitational potential, defined so that \(\Phi(r)\rightarrow0\) for \(r\rightarrow\infty\). To get the potential, we consider a mass distribution described by a singular isothermal sphere, with density \(\rho(r)=\rho_0r_0^2/r^2\), where \(\rho_0\) and \(r_0\) are constants. This density distribution generates a constant circular velocity \(v_\mathrm{c}=\sqrt{4\pi G\rho_0r_0^2}\), which represents the flat part of the rotation curve, such as the one in Figure \ref{fig:plo}(a). For a singular isothermal sphere, truncated at a maximum radius \(r_\mathrm{max}\) (needed to avoid mass divergence at large radii, because \(M(r)=4\pi\rho_0r_0^2r\)), the potential arising from the mass within \(r<r'<r_\mathrm{max}\) is \(\Phi(r'>r)=-4\pi G\rho_0r_0^2\ln(r_\mathrm{max}/r)=-v_\mathrm{c}^2\ln(r_\mathrm{max}/r)\). Since the mass distribution is assumed to be spherically symmetrical, the potential from the mass within \(0<r'<r\) is \(\Phi(r'<r)=-GM(<r)/r=-v_\mathrm{c}^2\).

The total potential at \(r\) is \(\Phi(r)=\Phi(r'<r)+\Phi(r'>r)\), and from Equation \ref{eq:esc} the escape velocity becomes

\begin{equation}
v_\mathrm{esc}(r)=v_\mathrm{c}\sqrt{2\left[1+\ln\left(\frac{r_\mathrm{max}}{r}\right)\right]}.
\end{equation}
Using this equation, and the rotation curve derived in section \ref{sec:rcm}, we get \(v_\mathrm{esc}(r)\approx250\mathrm{-}390\) km s\(^{-1}\) at \(r=30\farcs4\) (2.9 kpc) for a halo radius of \(r_\mathrm{max}=30\farcs4\mathrm{-}121\arcsec\), where the lower boundary is the radius within which CO is detected and rotation curve measured and we set the upper boundary to the \(3.4~\micron\) radius of stellar distribution \citep{Sor19}. Given that the measured molecular gas velocities are significantly lower than this, most of the molecular gas must be bound to the galaxy and eventually end up flowing as a galactic fountain. On the other hand, the velocity of the warm ionized gas outflow is \(v_\mathrm{w,H\alpha}\approx250~\mathrm{km~s^{-1}}\); a fraction of ionized gas in the high-velocity tail might flow into the CGM and escape into the IGM.

For an ionized gas mass outflow rate of \(\dot{M}_\mathrm{w,H\alpha}\sim0.6~M_\sun~\mathrm{yr}^{-1}\), the mass loss of NGC 1482 to the IGM may have been at most \(M_\mathrm{esc}\sim10^6~M_\sun\) over the period of the current starburst episode (assumed to be \(t_\mathrm{w}\approx10\) Myr). This is \(\sim10\%\) of the mass converted into stars during the same time interval if SFR is assumed constant. Compared to the mass of the molecular gas disk (\(M_\mathrm{mol}\sim2.7\times10^9~M_\sun\); section \ref{sec:mgd}), the mass lost to the IGM in the current starburst episode seems to be a modest fraction of the ISM and most of the ejected gas will resupply star formation in the process of recycled wind accretion (e.g., \citealt{Opp10,AA17}). With the current mass loss rate and SFR, the molecular gas reservoir in the galaxy will be depleted after \(t_\mathrm{dep}\sim7\times10^8\) yr.

Note that CO emission is highly concentrated toward the galactic center, as shown in Figure \ref{fig:plo}(b), meaning that the starburst-driven wind has not cleared the central region of molecular gas, and suggesting that the feedback may be in its early stage. There must be a mechanism that efficiently supplies molecular gas and maintains high SFR in the central 500 pc during the timescale of the superwind. In general, from the conservation of mass (Equation \ref{eq:ce}), the interstellar gas budget in the galactic center region is a sum of mass outflow rate, star formation rate, and mass inflow rate, where the last term includes external accretion, wind recycling, and gas returned to the ISM from the stars which is not loaded into the outflow.

From the CO (1--0) flux and kinematics of the molecular cloud denoted by H in Figure \ref{fig:pvd1}, we can probe the mass inflow rate in molecular phase from the halo. The inflow velocity is assumed to be equal to the difference between the observed velocity of the cloud H and the disk, \(v_\mathrm{in}\approx170\) km s\(^{-1}\) (see section \ref{sec:okin}). This is a simplistic but reasonable estimate, because the velocity is comparable within a factor of two to the current outflow velocities of molecular and ionized gases (100 and 250 km s\(^{-1}\), respectively), as expected if the clouds originate from the wind fountain. The total CO (1--0) flux in clouds H marked in Figure \ref{fig:pvd1} is \(\approx2\) Jy km s\(^{-1}\), and from equation \ref{eq:mol} we obtain a total mass of \(M_\mathrm{in}\approx2\times10^6~M_\sun\), where \(X_\mathrm{CO}=0.5\times10^{20}~\mathrm{cm^{-2}(K~km~s^{-1})^{-1}}\) was applied. From the distance to the disk (Figure \ref{fig:pvd2}), which we adopt to be \(\approx1.5\) kpc, the mass inflow rate becomes \(\dot{M}_\mathrm{in}\sim0.2~M_\sun~\mathrm{yr}^{-1}\). At present, the accretion in molecular phase appears to be inefficient, with \(\dot{M}_\mathrm{in}\ll\dot{M}_\mathrm{w}\).

It is also possible to make a simple estimate of the maximum distance \(r_\mathrm{f}\) that cloud H travelled from the disk before it halted and started falling back until it reached its present location \(r\) and inflow velocity \(v_\mathrm{in}(r)\). We assume ballistic motion and that mechanical energy is conserved, i.e., \(v^2_\mathrm{in}(r)/2=\Phi(r_\mathrm{f})-\Phi(r)\), and use the potential from above to find \(v_\mathrm{in}(r)=v_\mathrm{c}\sqrt{2\ln(r_\mathrm{f}/r)}\), which yields \(r_\mathrm{f}\approx2.4\) kpc for cloud H observed at \(r=1.5\) kpc with \(v_\mathrm{in}=170\) km s\(^{-1}\). This is well within the extent of the outflow, which is \(\sim5\) kpc, as found from X-ray and infrared observations (see the following section), suggesting that the fountain origin is possible.

\begin{figure*}
 \centering
  \includegraphics[width=1\textwidth]{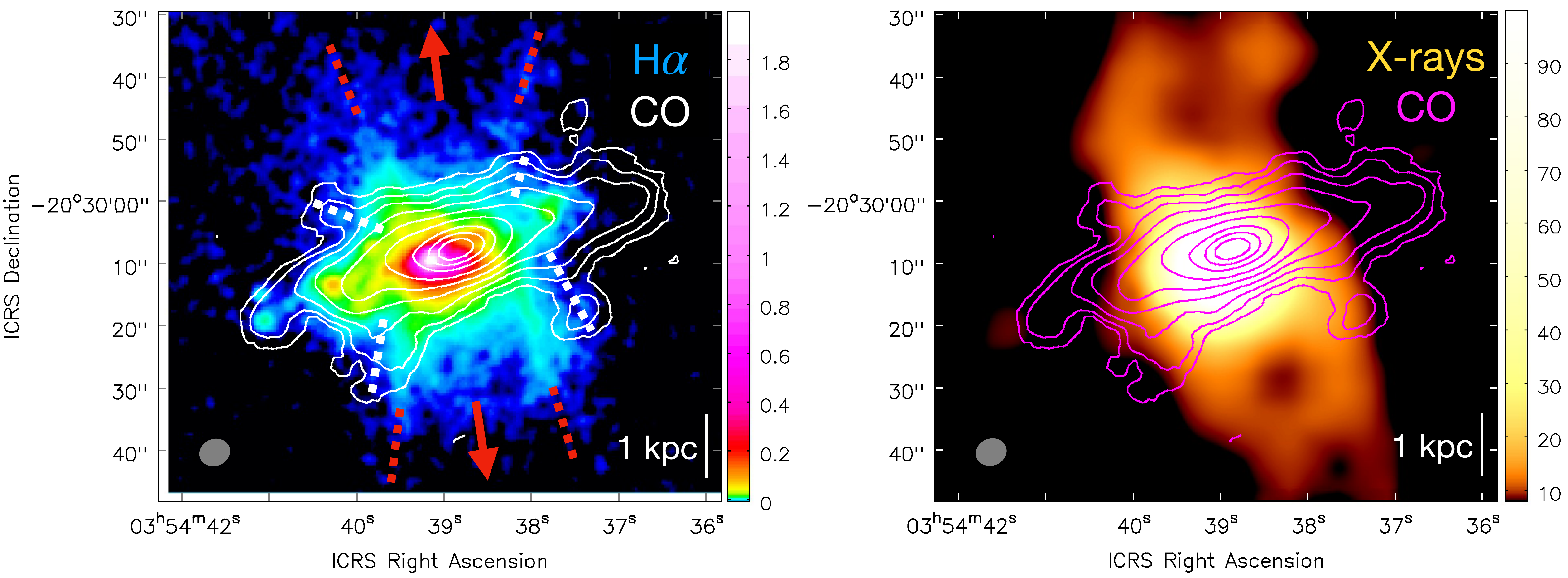}
 \caption{CO (1--0) integrated intensity map (\(uv\)-tapered) superimposed on H\(\alpha\) (left) and soft X-ray (0.3-2.0 keV) images (right). The contours are plotted at \((0.00625,0.0125,0.025,0.05,0.1,0.2,0.4,0.6,0.8)\times129.8~\mathrm{Jy~beam^{-1}~km~s^{-1}}\) (maximum). The image units are arbitrary. The dotted lines indicate ionized gas (red) and molecular gas (white) streamers and the arrows indicate the direction of the outflow.\label{fig:cohx}}
\end{figure*}

\subsection{Multiphase Outflow}\label{sec:mul}

Multiphase medium is ubiquitous in starburst-driven outflows. Although there has been significant progress in recent years in understanding the physics of the winds, some key properties, such as the gas conditions (density and temperature), launching mechanism, entrainment of cold gas and dust in a hot wind, and the survival of cold clouds are not fully understood (e.g., \citealt{Coo09,Fuj09,McC15,SB15,Tho16,SR17,GO18,SRT18,KKW20}). To study such complex media, it is important to reveal the spatial distribution of each ISM phase including cold molecular gas. In this section, we compare the distributions of CO, H\(\alpha\), and soft X-rays in the wind of NGC 1482, as representatives of cold neutral, warm ionized, and hot ionized gas phases, respectively. We also compare CO with dust continuum at \(70~\micron\) and \(100~\micron\).

In order to maximize sensitivity to large-scale structure, we created a CO (1--0) data cube by applying a taper (Gaussian of FWHM \(4\arcsec\)) on visibilities in the \emph{uv} plane in combination with natural weighting. The FWHM beam size of the data cube that resulted from this process is \(b_\mathrm{maj}\times b_\mathrm{min}=4\farcs902\times4\farcs219\) (position angle \(-68\fdg9\)) and the sensitivity is \(4.1~\mathrm{mJy~beam^{-1}}\) (equivalent to 18 mK) at a spectral resolution of 10.16 km s\(^{-1}\). In Figure \ref{fig:cohx}, we show a CO (1--0) integrated intensity map created from the tapered cube. The integrated intensity map was created by integrating the intensity over velocity after masking the cube at \(2\sigma\).

\begin{figure}
 \centering
  \includegraphics[width=0.475\textwidth]{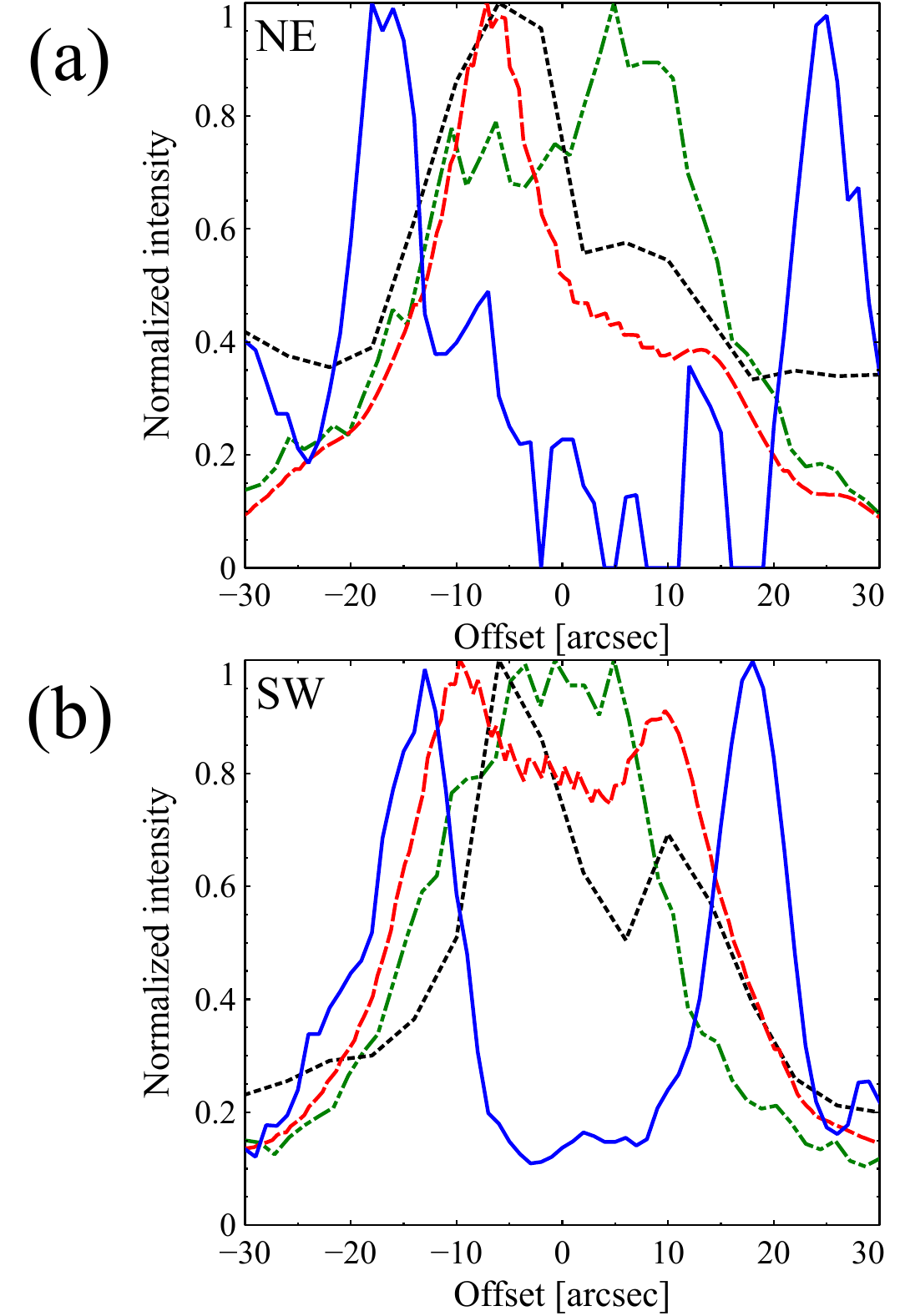}
 \caption{ISM phases in the wind: normalized intensity profiles at (a) \(Z=+15\arcsec\) NE and (b) \(Z=-15\arcsec\) SW of the galaxy center, where \(Z\) is the galactic minor axis (\(Z=z\sin{i}\)). The offset is parallel to the galactic major axis increasing from SE to NW. The curves are CO (solid blue), H\(\alpha\) (dashed red), soft X-rays (dotted black), and \(70~\micron\) (dash-dotted green) smoothed to the same resolution of \(5\farcs6\).\label{fig:cut}}
\end{figure}

\subsubsection{Warm Ionized Gas}

The CO (1--0) integrated intensity map in Figure \ref{fig:cohx} (left panel) is superimposed on a continuum-subtracted H\(\alpha\) image, acquired by the Cerro Tololo Inter-American Observatory (CTIO) 1.5 m telescope, that traces warm (\(T\sim10^4\) K) ionized gas \citep{Ken03}. The comparison shows that CO is distributed up to \(h_z\approx2\) kpc above the galactic plane with a similar biconical, hourglass symmetry as the H\(\alpha\) wind. However, note that CO streamers are generally offset from those of H\(\alpha\) at this scale, with a projected separation of 4\arcsec--10\arcsec (0.4--1 kpc) between the intensity peaks. The offset between CO and H\(\alpha\) for major streamers is also clearly visible in the intensity profiles at the projected distance of \(|Z|=15\arcsec\) (1.5 kpc) north and south of the galactic disk shown in Figure \ref{fig:cut}. The cold CO gas occupies layers at larger radii from the wind axis compared to the warm ionized gas, possibly reflecting temperature, ionization, and density gradients. The distributions in Figure \ref{fig:cohx} (left panel) suggest that at least on the south side of the outflow, which is closer to the observer, the offset exists already close to the starburst disk (\(z\lesssim0.5\) kpc), where the outflow is launched.

\subsubsection{Hot Ionized Gas}

The superwind is also glowing in X-rays. Figure \ref{fig:cohx} (right panel) shows a soft X-ray image, acquired by the \emph{Chandra} telescope, that traces plasma gas at a temperature of \(T\sim10^6~\mathrm{K}\) \citep{Str04a,Str04b}. The distribution of X-ray-emitting gas is elongated and extends to 5 kpc perpendicular to the molecular gas disk, exhibiting a biconical structure with a depression in the central region of the wind along the \(z\)-axis. At this angular resolution, X-rays appear to arise from regions that are also bright in H\(\alpha\), where shock ionization is taking place \citep{VR02}. Figure \ref{fig:cut} shows that the X-ray-emitting wind and H\(\alpha\) exhibit generally similar distributions in most streamers, albeit H\(\alpha\) is distributed more outward in some places, such as the region at offset \(-10\arcsec\) on the SW side where H\(\alpha\) occupies the space between CO and X-rays and the phases exhibit maxima at mutual separations of \(\approx5\arcsec\) (\(\approx500\) pc). This transition is consistent with the picture of a temperature gradient across the outflow cones. The central part of the biconical wind, where a depression is observed in all three tracers, may be filled with a very hot (\(T\gtrsim10^7\) K), tenuous plasma (wind fluid), which is a weak X-ray emitter \citep{SS00}. A similar biconical soft X-ray halo and its coincidence with H\(\alpha\) is also reported for NGC 253 \citep{Str02}, indicating that the observed radiation may originate from the mixing of hot (\(\sim10^6\) K) and warm (\(\sim10^4\) K) phases. \citet{Coo09} suggested that bow shocks that form upstream of dense clouds entrained in a hot wind and their interactions are responsible for the soft X-ray emission. On the other hand, molecular gas (\(\lesssim10^2\) K) in NGC 1482, traced by CO (1--0) emission, is clearly distributed farther out from the wind axis and forms an envelope around the hot wind. The molecular gas has probably been transported from the starburst disk, but some of it may have recombined \emph{in situ} from shocked ionized gas that has been cooling in the outer layer of the expanding hot wind. Reformation of molecules may be possible in starburst-driven expanding shells \citep{Roy16}, and the presence of cool dust grains in the outflow could make this process efficient owing to their catalyzer role in H\(_2\) formation \citep{HS71}. However, molecular gas entrained in the wind is exposed to shocks from the expanding hot gas and may also be dissociated in the outflow. In that case, H\(_2\) gas undergoes a transition into \ion{H}{1} and \ion{H}{2} and may add material to the warm gas phase. Indeed, clouds of cold gas are expected to be disrupted when embedded in flows of hot and fast wind gas (e.g., \citealt{SB15}). Observations of neutral atomic gas at comparable resolution may help us clarify the phase transitions in the outflow.

\begin{figure*}
 \centering
  \includegraphics[width=1\textwidth]{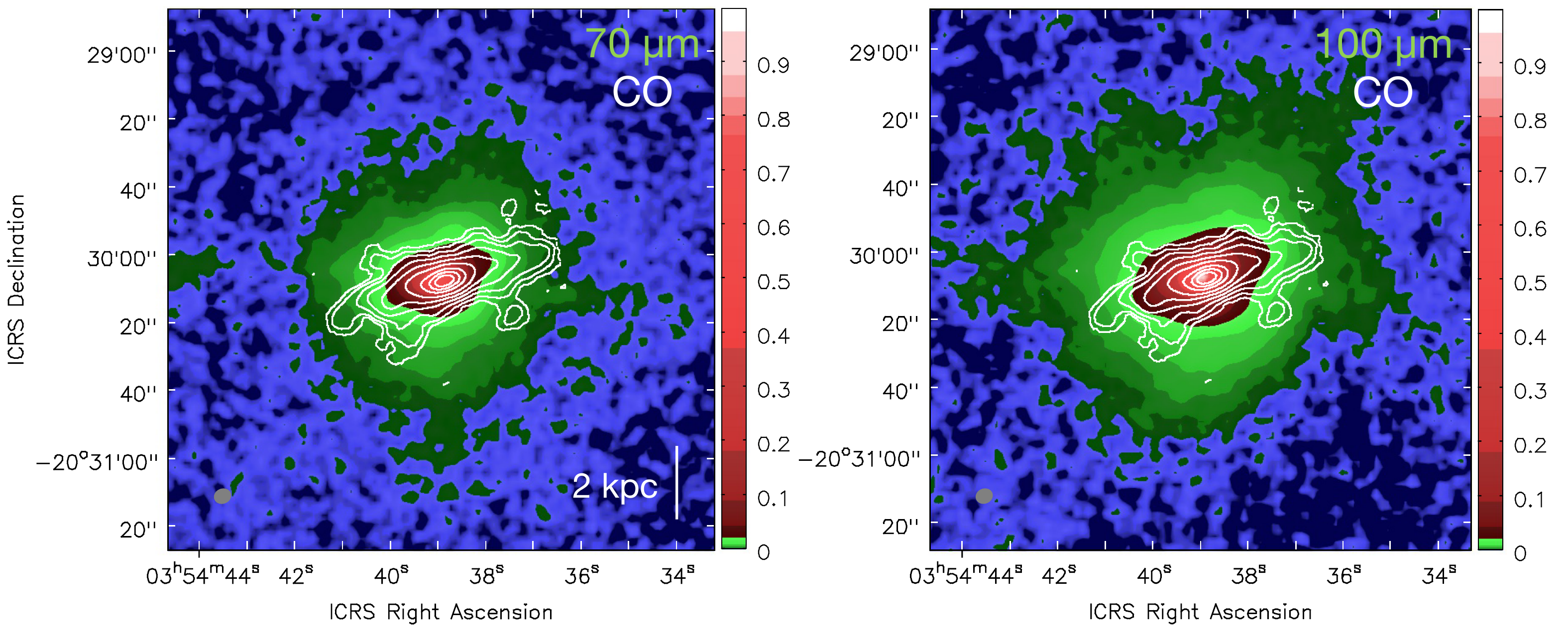}
 \caption{CO (1--0) integrated intensity map (contours as in Figure \ref{fig:cohx}) superimposed on \(70~\micron\) (left) and \(100~\micron\) (right) continuum images (color) with intensity normalized to unity.\label{fig:FIR}}
\end{figure*}

\subsubsection{Dust}

In Figure \ref{fig:FIR}, we compare CO distribution with that of \(\lambda=70~\micron\) and \(100~\micron\) dust continuum imaged by the PACS instrument on \emph{Herschel} \citep{Ken11}. These data are at a resolution of \(5\farcs6\) and \(6\farcs8\), respectively, which is comparable to that of CO (\(uv\)-tapered). Interestingly, unlike the gas phases discussed above, the FIR-emitting dust is distributed in an approximately spherical halo that extends up to 5 kpc from the galactic center with no clear correlation with any of the gas phases. Moreover, figure \ref{fig:FIR} shows that the \(100~\micron\) continuum is more extended than that of \(70~\micron\), indicating a change in dust properties (e.g., temperature) in the halo. At first glance, the dust appears to be anti-correlated with CO and H\(\alpha\) in the outflow. The anti-correlation with H\(\alpha\), which is most prominent in the NE panel of Figure \ref{fig:cut}, can occur in the case that extraplanar dust is absorbing the background H\(\alpha\) photons, but the dust here is not associated with CO. On the other hand, the \(70~\micron\) continuum maximum is in the middle between the CO maxima, indicating that the continuum may be preferentially tracing warm dust entrained in the wind. Since dust is detected up to 5 kpc in the halo, which is comparable to the spatial extent of the X-ray emitting gas, the result implies that grains can survive in the wind for \(\sim1\times10^7\) yr, the dynamical timescale of the gas outflow.

The distribution of \(100~\micron\) continuum north of the center appears to be conical, with two streamers emerging from the central region. However, the dust is also distributed widely outside of the biconical gaseous outflow, suggesting that grains are entrained in both warm and cool phases in the wind and halo. A similar spatial distribution is also observed in PAH molecules detected at \(7.7~\micron\) by the IRAC instrument on \emph{Spitzer} \citep{MVR13}. These authors reported filamentary structure of PAH emission extending in many directions. Similar results were also found in M82, where an approximately spherical dust halo is observed on a scale of several kiloparsecs with evidence that dust grains and PAHs are entrained in both neutral and ionized gas phases (e.g., \citealt{Kan10,Rou10}).

\subsubsection{Multiphase Outflow Summary}

The observations of the various ISM gas phases discussed above are summarized in Figure \ref{fig:outm}. In our outflow model, very hot (\(\gtrsim10^7\) K), low-density plasma is flowing out of the starburst disk along the minor galactic axis. It is surrounded by hot (\(\sim10^{6\mathrm{-}7}\) K) and warm (\(\sim10^4\) K) ionized gas emitting soft X-rays and H\(\alpha\), respectively. Shock ionization is taking place in the warm gas giving rise to the high [\ion{N}{2}]/H\(\alpha\) ratio of \(\gtrsim1\). Farther out, gas density is high and temperature low (\(\lesssim10^2\) K) enough for the ISM to be in molecular phase. This region is observed in CO (1--0), but may also contain \ion{H}{1} at the interface between CO and H\(\alpha\). The dust traced by FIR thermal continuum is distributed throughout the multiphase wind and halo up to a distance of 5 kpc from to the galactic center, comparable to the spatial extent of the hot gas.

\subsection{Tidal Interaction and the Origin of the Starburst}\label{sec:ori}

\begin{figure}
 \centering
  \includegraphics[width=0.475\textwidth]{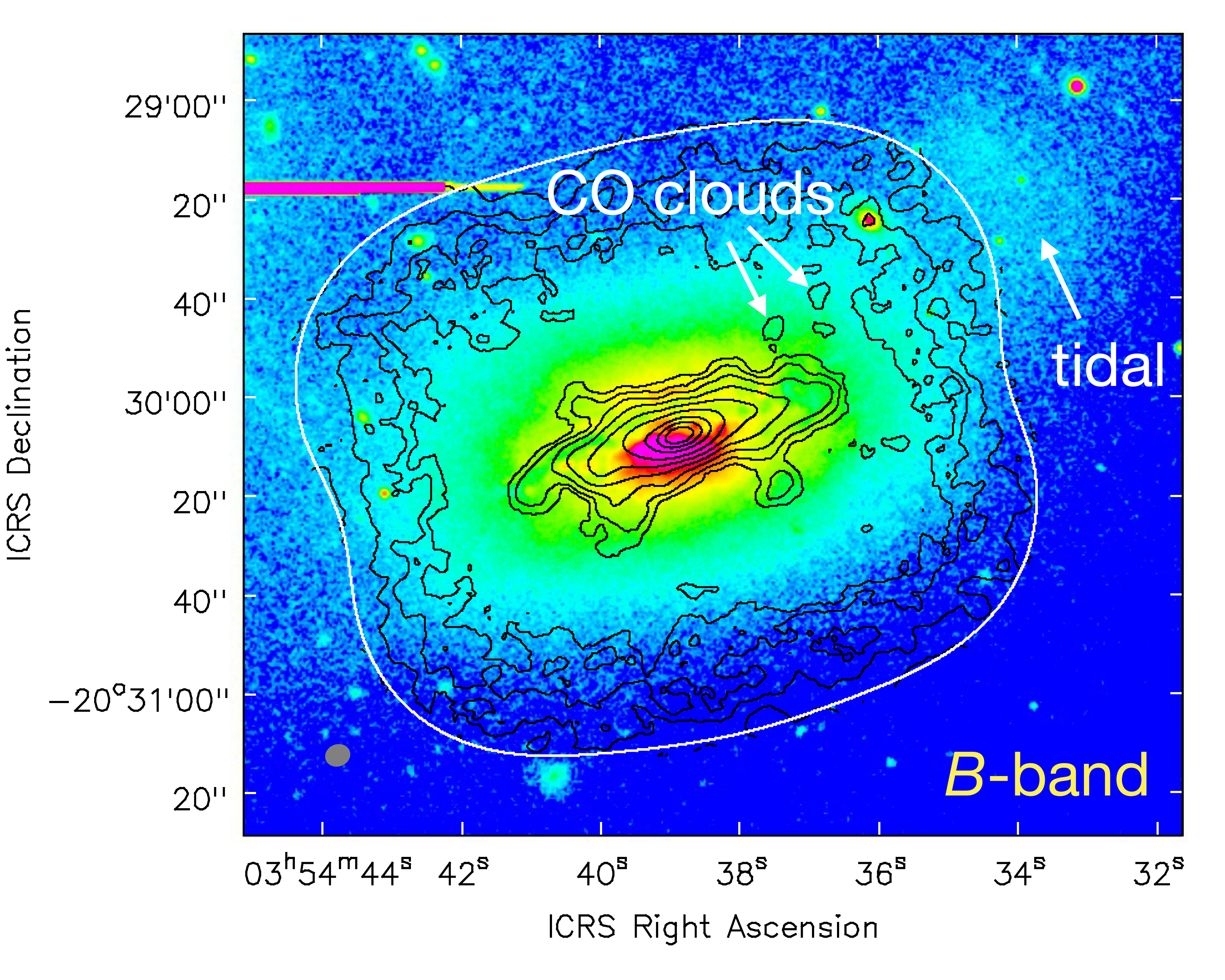}
 \caption{CO (1--0) integrated intensity map (contours as in Figure \ref{fig:cohx}) superimposed on a false-color \(B\)-band image \citep{Ken03}. CO clouds in the halo (the H component in Figure \ref{fig:pvd1}) and a stellar tidal feature are indicated by arrows. The white contour marks the total region observed by ALMA. The CO map has higher rms noise at the edge because of reduced primary beam response.\label{fig:tid}}
\end{figure}

How did NGC 1482, an early-type galaxy, acquire so much molecular gas to fuel an X-ray glowing superwind? In section \ref{sec:okin}, the H feature in pv diagrams (Figures \ref{fig:pvd1} and \ref{fig:pvd2}) was discussed in the context of possible tidal interaction between NGC 1482 and its neighbor NGC 1481 in the Eridanus group of galaxies. In Figure \ref{fig:tid} we show a CTIO \(B\)-band optical image of NGC 1482 \citep{Ken03} including CO (1--0) contours of the \(uv\)-tapered map. The distribution of stars forms the shape of an early-type galaxy. However, note a blob of stars at a projected distance of \(\approx70\arcsec\) (6.7 kpc) northwest of the center of the galaxy. This feature has also been found at 3.6 \(\micron\) that traces old stellar population \citep{Kim12} and indicates a tidal interaction. We also note that there are molecular clouds in the halo approximately in the direction toward the blob. The velocity of the clouds is comparable to the systemic velocity of NGC 1482 which suggests motion in the plane of the sky. Although the clouds could also be ejected from the superwind, tidal origin cannot be ruled out.

Tidal interactions between galaxies provide an efficient mechanism to transport cold gas toward galactic central regions. A particularly impressive example is the M81-M82 system. M82 has been closely interacting with its neighbor M81 and the two galaxies are immersed in a giant halo of \ion{H}{1} gas \citep{YHL93,deB18}. This gas reservoir may have supplied M82's starburst. Similarly, NGC 1482 has been interacting with its neighbor NGC 1481. \citet{OD05} reported an extended \ion{H}{1} reservoir with prominent tidal tails in the NGC 1481-1482 system. Since the disk in NGC 1482 appears to be connected to this large-scale \ion{H}{1} structure, we suggest that gas fueling, followed by the starburst and superwind in NGC 1482 were triggered by a tidal interaction. In the context of a baryon cycle, the current starburst is fueled by external accretion from the intergalactic \ion{H}{1} gas reservoir, whereas future star formation will also be maintained by re-accretion of the gas ejected by the wind.

\section{Summary}\label{sec:sum}

In this article, we have presented the first high-resolution observations of molecular gas traced by CO (1--0) and the discovery of a molecular outflow in the nearby starburst galaxy NGC 1482. The main results of the study are summarized below.

\begin{enumerate}

\item The observations of CO (1--0) and 100 GHz continuum were carried out by ALMA in cycle 7 at a resolution of 1\arcsec (\(\approx100\) pc). We conducted mosaic observations over the central \(r<4\) kpc region using the 12 m and 7 m (ACA) arrays.

\item Molecular gas is distributed in a nearly edge-on disk (inclination \(i=76\arcdeg\)) with a radius of 3 kpc and outflow (streamers) extending at least 1.5 kpc perpendicular to the disk. This is the first detection of a molecular wind in this galaxy. The 100 GHz continuum is detected in the central 1 kpc, which is the base of the outflow. From the continuum flux density, we derived a star formation rate of \(\mathrm{SFR}\approx4~M_\sun\) yr\(^{-1}\) in the central starburst region.

\item We modeled the 3-kpc disk using the \(^\mathrm{3D}\)Barolo program. Basic parameters of the disk, including the position angle, inclination, rotational velocity, and mass surface density of molecular gas in the central \(R<3\) kpc were derived. For a standard Galactic CO-to-H\(_2\) conversion factor \(X_\mathrm{CO}=2\times10^{20}~\mathrm{cm^{-2}(K~km~s^{-1})^{-1}}\), the total molecular gas mass (\(M_\mathrm{mol}=2.7\times10^9~M_\sun\)) was found to be 12\% of the dynamical mass within the central \(R<3\) kpc.

\item Using the disk model, we estimated that the CO (1--0) flux originating from the extraplanar molecular outflow is \(\approx10\%\) of the total flux, yielding an outflow mass of \(M_\mathrm{w}\sim7\times10^7~M_\sun\), where a low CO-to-H\(_2\) conversion factor of \(X_\mathrm{CO}=0.5\times10^{20}~\mathrm{cm^{-2}(K~km~s^{-1})^{-1}}\) was applied. The base of the molecular outflow has a radius of \(R\approx1\) kpc. We also estimated the molecular wind velocity \(v_\mathrm{w}\sim100\) km s\(^{-1}\) from the kinematics of the extraplanar gas. The mass outflow rate is \(\dot{M}_\mathrm{w}\sim7~M_\sun~\mathrm{yr^{-1}}\), and the mass loading factor of the molecular wind is \(\dot{M}/\mathrm{SFR}\sim2\). These results were used to calculate the kinetic energy \(E_\mathrm{w}\) and momentum \(p_\mathrm{w}\) of the wind. We found that \(E_\mathrm{w}\) and \(p_\mathrm{w}\) are, respectively, \(\sim1\%\) and \(\sim20\%\) of the initial energy and momentum released by supernova explosions in the central 1 kpc. Considering that there is no evidence of an AGN, all results suggest that the superwind is driven by starburst feedback.

\item The molecular outflow velocity is significantly lower than the escape velocity \(v_\mathrm{esc}\) at a radius of 3 kpc. Therefore, most of molecular gas is bound to the galaxy and is likely to fall back onto the disk and resupply star formation. On the other hand, the velocity of warm ionized gas outflow is comparable to \(v_\mathrm{esc}\); a fraction of this gas may be escaping into the IGM. During the current starburst episode (\(\sim1\times10^7\) yr), if SFR has been constant, NGC 1482 may have lost \(M_\mathrm{esc}\sim10^6~M_\sun\) of its interstellar gas, which is a small fraction (\(10^{-3}\)) of the molecular gas reservoir. At the current SFR and mass outflow rate, the depletion time of molecular gas is \(t_\mathrm{dep}\sim7\times10^8~\mathrm{yr}\). The mass inflow rate in molecular phase from the halo is estimated to be \(\dot{M}_\mathrm{in}\sim0.2~M_\sun~\mathrm{yr}^{-1}\), indicating that accretion is inefficient (\(\dot{M}_\mathrm{in}\ll\dot{M}_\mathrm{w}\)) in the current evolutionary stage of the wind.

\item We present a model of a multiphase, cylindrically symmetrical, hourglass-shaped superwind. Very hot (\(T\gtrsim10^7\) K), tenuous plasma gas occupies the central region along the wind axis, surrounded by hot (\(\sim10^6\) K) and warm (\(\sim10^4\) K) ionized gas phases, enveloped by cold (\(\lesssim10^2\) K) gas traced by CO (1--0). The multilayer model is supported by observations of soft X-rays, H\(\alpha\), and CO, which reveal that cold and warm gases are clearly separated across the cylindrically symmetrical wind at a distance of 1 kpc above the galactic plane. We found that the dust continuum at \(70~\micron\) and \(100~\micron\) is distributed throughout the wind and halo up to a distance of 5 kpc perpendicular to the galactic disk and exhibits no clear spatial correlations with these gas phases.

\item There is evidence that NGC 1482 has experienced a tidal interaction with its neighbor NGC 1481. In particular, there are tidal tails observed in stellar and atomic gas (\ion{H}{1}) distributions. We suggest that the starburst and superwind in NGC 1482 were triggered by tidal interaction that led to a rapid supply of neutral gas into the galactic central region.

\end{enumerate}

\acknowledgments

The authors thank the referee for many comments and suggestions that improved the manuscript. This paper makes use of the following ALMA data: ADS/JAO.ALMA\#2019.1.01132.S. ALMA
is a partnership of ESO (representing its member states), NSF (USA) and NINS (Japan),
together with NRC (Canada), MOST and ASIAA (Taiwan), and KASI (Republic of Korea), in
cooperation with the Republic of Chile. The Joint ALMA Observatory is operated by
ESO, AUI/NRAO and NAOJ. Based on observations made with the NASA/ESA Hubble Space Telescope, and obtained from the Hubble Legacy Archive, which is a collaboration between the Space Telescope Science Institute (STScI/NASA), the Space Telescope European Coordinating Facility (ST-ECF/ESA) and the Canadian Astronomy Data Centre (CADC/NRC/CSA). This publication makes use of data products from the Two Micron All Sky Survey, which is a joint project of the University of Massachusetts and the Infrared Processing and Analysis Center/California Institute of Technology, funded by the National Aeronautics and Space Administration and the National Science Foundation. This research has made use of data obtained from the Chandra Data Archive and the Chandra Source Catalog, and software provided by the Chandra X-ray Center (CXC) in the application packages CIAO, ChIPS, and Sherpa. This research has made use of the NASA/IPAC Extragalactic Database (NED), which is funded by the National Aeronautics and Space Administration and operated by the California Institute of Technology.

\end{document}